\documentclass[review]{elsarticle}

\usepackage[bookmarks,bookmarksnumbered]{hyperref}

\hypersetup{pdfauthor=author}
\usepackage{tabularx}
\usepackage{lineno}
\modulolinenumbers[5]
\usepackage{adjustbox}
\usepackage{amsmath,bm}
\usepackage{amssymb}
\usepackage{siunitx}
\usepackage{textcomp}
\usepackage{mathtools}
\usepackage{tabularx,multirow,makecell}
\usepackage[ruled,vlined]{algorithm2e}
\usepackage{graphicx,multirow}
\usepackage{booktabs}

\usepackage[flushleft]{threeparttable}

\usepackage{setspace}
\usepackage{xcolor}
\usepackage{color}
\usepackage{soul}
\usepackage{rotating}

\def\XS{\xspace}

\def\etal{\textit{et al.}\XS}

\def\ie{\textit{i.e.,}\xspace}

\DeclareMathAlphabet{\mathb}{OML}{cmm}{b}{it}
\def\sbm#1{\ensuremath{\mathb{#1}}}                % Style gras italique (necessite amsmath)
\def\scu#1{\ensuremath{\mathcal{#1\XS}}}           % Style cursif
\def\Sb{{\sbm{S}}\XS}  
\def\Xb{{\sbm{X}}\XS}  
  \def\pb{{\sbm{p}}\XS}
\def\Mb{{\sbm{M}}\XS}  \def\mb{{\sbm{m}}\XS}
\def\Ab{{\sbm{A}}\XS}  
  \def\zb{{\sbm{z}}\XS}
\def\Eb{{\sbm{E}}\XS}  
\def\Wb{{\sbm{W}}\XS}

\def\Nc{{\scu{N}}\XS}   
\def\Ac{{\scu{A}}\XS}   
\def\Lc{{\scu{L}}\XS}   
\def\Xc{{\scu{X}}\XS}

\definecolor{brickcolor}{rgb}{0.98, 0.32, 0.15}
\definecolor{seacolor}{rgb}{0.1796875, 0.64296875, 0.83984375}

\usepackage{float}
\usepackage{placeins}

\usepackage{pifont}
\newcommand{\xmark}{\ding{55}}%

\usepackage{setspace}
\journal{Neurocomputing}

\begin{document}

\begin{frontmatter}
\title{Leveraging Statistical Shape Priors in GAN-based ECG Synthesis}
%\title{Elsevier \LaTeX\ template\tnoteref{mytitlenote}}

\author[redcad]{Nour~Neifar\corref{mycorrespondingauthor}}
\ead{nour.neifar@redcad.org}
\author[crns]{Achraf~Ben-Hamadou}\ead{achraf.benhamadou@crns.rnrt.tn} 
\author[redcad]{Afef~Mdhaffar} 
\ead{afef.mdhaffar@enis.tn}
\author[redcad]{Mohamed~Jmaiel} 
\ead{mohamed.jmaiel@redcad.org}
\author[marburg]{Bernd~Freisleben}
\ead{freisleben@uni-marburg.de}

\address[redcad]{ReDCAD Lab, ENIS, University of Sfax, Tunisia}
\address[crns]{Centre de Recherche en Num\'{e}rique de Sfax, Laboratory of Signals, Systems, Artificial Intelligence and Networks, Technopôle de Sfax, Sfax, Tunisia}
\address[marburg]{Department of Mathematics and Computer Science, Philipps-Universität Marburg, Germany}
\cortext[mycorrespondingauthor]{Corresponding author}

\begin{abstract}
Electrocardiogram (ECG) data collection during emergency situations is challenging, making ECG data generation an efficient solution for dealing with highly imbalanced ECG training datasets. In this paper, we propose a novel approach for ECG signal generation using Generative Adversarial Networks (GANs) and statistical ECG data modeling. Our approach leverages prior knowledge about ECG dynamics to synthesize realistic signals, addressing the complex dynamics of ECG signals. To validate our approach, we conducted experiments using ECG signals from the MIT-BIH arrhythmia database. Our results demonstrate that our approach, which models temporal and amplitude variations of ECG signals as 2-D shapes, generates more realistic signals compared to state-of-the-art GAN based generation baselines. Our proposed approach has significant implications for improving the quality of ECG training datasets, which can ultimately lead to better performance of ECG classification algorithms. This research contributes to the development of more efficient and accurate methods for ECG analysis, which can aid in the diagnosis and treatment of cardiac diseases.
\end{abstract}


\end{frontmatter}

%\linenumbers

\section{Introduction\label{sec:intro}}
\begin{figure}[t]
\centering
\includegraphics[width=0.5\textwidth]{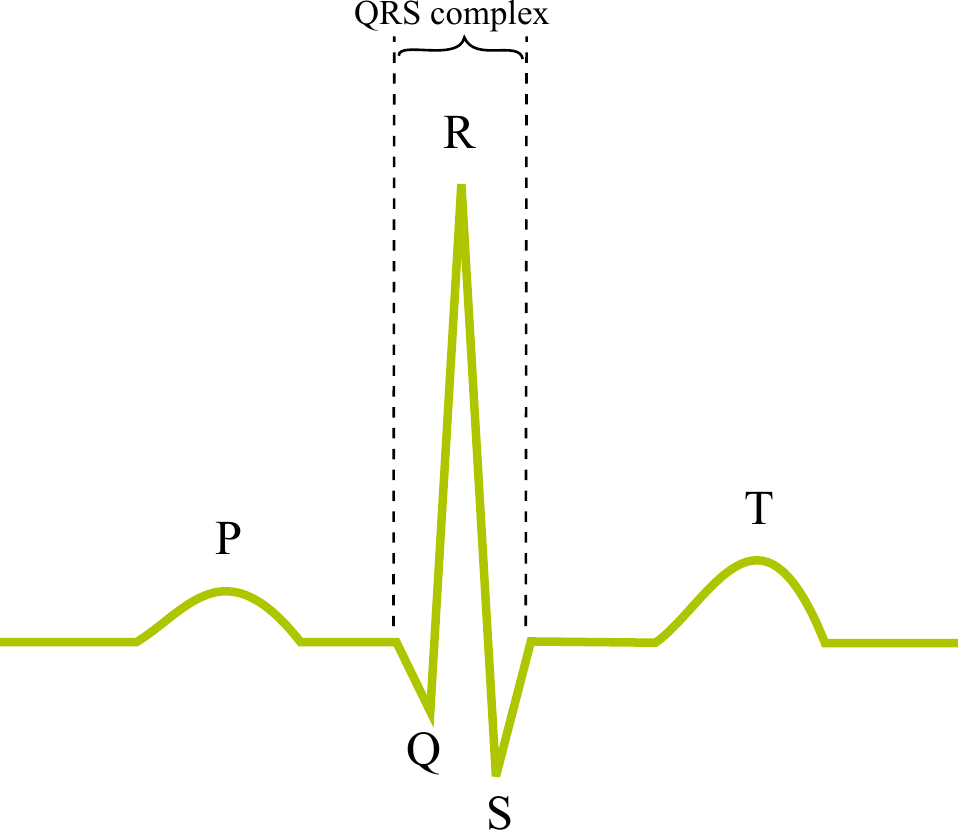}
\caption{Illustration of  ECG waves including the P wave followed by the QRS complex and the T wave.}
\label{ecg}
\end{figure}

Deep learning has been successfully applied in a wide range of fields, such as natural language processing, computer vision, and e-health systems \cite{ Otterasurvey,WuTheapplicationof,ZHAO2019213}. Nevertheless, one of the persistent challenges in these fields, particularly in medicine, is the scarcity of training data, primarily caused by ethical and legal regulations \cite{gerke2020ethical}. Furthermore, collecting pathological data during emergencies, such as heart attacks or epileptic seizures, is extremely difficult, resulting in highly imbalanced datasets. Consequently, various techniques for medical data synthesis have recently emerged in an attempt to address these challenges. 
%In particular, Generative Adversarial Networks (GANs) \cite{goodfellow2014generative} have shown their efficiency in generating high quality images (\eg facial images \cite{zhang2020high}, medical images \cite{skandarani2021gans,shin2018medical}), videos \cite{tulyakov2018mocogan}, and time series \cite{yoon2019time,smith2020conditional,smith2021spectral}) 
In particular, Generative Adversarial Networks (GANs) have shown their efficiency in generating high-quality facial images \cite{zhang2020high}, medical imaging  \cite{skandarani2023gans,shin2018medical}, videos \cite{tulyakov2018mocogan}, and time series such as electrocardiograms (ECGs) \cite{yoon2019time,smith2020conditional,smith2021spectral}.
In generating ECGs, one of the key challenges is capturing the complex dynamics of ECG signals. ECG signals are time series of physiological significance generated by recording electrical activities of the heart over time. They are characterized by a sequence of heartbeats, each of which corresponds to a cardiac cycle represented by a series of electrical and mechanical events. A normal ECG signal exhibits distinct patterns for each event, including a P wave, a QRS complex, and a T wave, each with its own unique shape and timing (see Figure~\ref{ecg}). The morphology and timing of ECG signals can vary significantly across individuals, making it challenging to generate realistic ECG signals that accurately capture the dynamics of physiological variations. Moreover, generating realistic ECG signals requires addressing several additional challenges, such as dealing with noise and artifacts in the ECG signal, handling variations in lead placement, and accounting for the effects of different body positions on the signal. These challenges are further complicated by the fact that pathological ECG signals can exhibit complex and irregular patterns, making it difficult to generate realistic synthetic data for training ECG classification models. These challenges underscore the importance of developing a more sophisticated approach to ECG signal generation that integrates prior knowledge of ECG dynamics and morphology, aiming to produce realistic and diverse ECG signals.
The main focus of this study is to leverage prior knowledge of ECG signal properties in order to generate accurate and realistic ECG waveforms. For this purpose, we integrate the capabilities of generative adversarial networks (GANs) with statistical shape modeling of ECG signals, enabling better control over the generation process.

Previous approaches based on deep learning models to generate synthetic ECG data \cite{delaney2019synthesis,zhu2019electrocardiogram,brophy2020synthesis} typically employ standard GAN architectures, which, however, fail to consider the dynamic properties inherent in complex physiological signals such as ECG. Several recent studies \cite{golany2019pgans,golany2020simgans,golany2021ecg,nour2021Disentangling} have made the efforts to address this challenge by integrating customized prior knowledge of ECG dynamics and patterns into the generation process, such as localization and order of ECG signal peaks. However, all of these features in \cite{golany2019pgans,golany2020simgans,golany2021ecg} are considered to be handcrafted. Indeed, they are based on domain-specific knowledge of ECG signals. 
Different from the approaches in \cite{golany2019pgans, golany2020simgans, golany2021ecg}, our previous solution presented in \cite{nour2021Disentangling} aimed to learn prior knowledge by utilizing ECG shape representatives and modeling the 1-D pattern dynamics of ECGs by disentangling the temporal and amplitude variations of the signal. However, it should be noted that the dissociation between temporal and amplitude variations may not be as sufficient and effective in accurately modeling the complex patterns and variations.

In this paper, we present a novel approach that incorporates statistical prior knowledge of ECG signal dynamics into the generation process using GANs. This allows us to efficiently control the generation process and produce more realistic and diverse ECG signals. We designed a set of statistical 2-D shape models that incorporate prior knowledge about the general shape of main ECG signal clusters as well as the variability within signals belonging to the same cluster. Our experimental results demonstrate the advantages of including statistical shape modeling in the generation process and the robustness of our method for generating ECG signals with realistic morphology. The proposed approach has significantly improved the performance of ECG classification algorithms, which can ultimately benefit patients by assisting in the diagnosis and treatment of cardiac diseases.

The remaining of the paper is structured as follows: Section \ref{sec:related_work} discusses the related work, Section \ref{sec:proposed_approach} presents our proposed approach, Section \ref{sec:experiments_and_results} describes the experimental evaluation and the obtained results, and finally, Section \ref{sec:conclusion} summarizes our findings and outlines potential future research directions.

\section{Related Work\label{sec:related_work}}
In this section, we describe the principles of GANs and provide an overview of existing methods for generating ECG signals. GANs are powerful tools for data generation. They consist of two neural networks: a generator network and a discriminator network. Taking a noise vector as its input, the generator attempts to generate data that are similar to real data. The discriminator, on the other hand, attempts to distinguish between real and fake data (\ie the output of the generator). The training process is defined as a competition between these two networks. In other words, the discriminator's goal is to learn to distinguish between real and fake data, whereas the generator's goal is to fool the discriminator.

Several GAN-based methods for ECG signal synthesis have been proposed in the literature. They can be divided into two categories. The first category includes methods based on the adaptation of standard GANs. The second category of methods focuses on integrating prior knowledge of ECG signal dynamics into deep generative models in order to improve their ability to generate realistic ECG data.

\subsection{ECG Generation Using Standard GANs}
Wang \etal \cite{wang2020accurate} proposed a simplified GAN architecture based on fully connected (FC) layers for generating ECG heartbeat data. The authors qualitatively validated their models using a loss graph generated during the training process and by a visual comparison between the generated ECG heartbeat data and real ECG heartbeat data.

Antczak \cite{antczak2020generative} presented an ECG generation method called ECG-GAN. The authors made few adjustments to the architecture of the deep convolutional generative adversarial network (DCGAN). To validate their approach, the generated data are added to real data to train an auto-encoder designed for filtering noise in the ECG signals.

Nankani \etal \cite{nankani2020investigating} proposed a conditional DCGAN model to synthesize different heartbeat classes. The synthesized heartbeats were quantitatively evaluated using various statistical metrics such as maximum mean discrepancy (MMD) and dynamic time warping (DTW).

Delaney \etal \cite{delaney2019synthesis} designed different GAN models combining convolutional neural networks (CNN) and Long Short-Term Memory (LSTM) in order to generate time series data like ECG.  The authors computed the MMD and DTW metrics after each training epoch to quantitatively evaluate the proposed models.

Zhu \etal \cite{zhu2019electrocardiogram} introduced a BiLSTM-CNN GAN framework for ECG generation. Three criteria were considered to evaluate the generated ECG signals: Fréchet distance (FD), percent root mean square difference (PRD) and root mean squared error (RMSE).

Hazra \etal \cite{hazra2020synsiggan} developed SynSigGan, a framework for generating biomedical signals including ECG. It is based on a bidirectional grid LSTM (BiGridLSTM) for the generator network and a CNN for the discriminator network. The authors evaluated the quality of the generated synthetic signals using various metrics such as the Mean Absolute Error (MAE), RMSE, PRD, and FD scores. They assessed the correlation between real signals and generated synthetic signals by calculating the Pearson Correlation Coefficient (PCC). Their findings revealed that the synthetic data are highly correlated with the original data, and combining BiGridLSTM-CNN with a GAN produces better results.

Brophy \etal \cite{brophy2020synthesis} presented a Multivariate GAN to synthesize physiological, dependent and multivariate time series data. The generator and discriminator networks are based on LSTM and CNN layers, respectively. The multivariate dependent DTW ($DTW_D$) was proposed and computed at the end of each epoch to evaluate the generated data, in addition to the MMD metric. The ($DTW_D$) was used to compare the similarity between the dependent multi-channel real data and the generated data.

Later, Brophy \etal \cite{brophy2021multivariate} improved their previous work by developing a GAN for the generation of dependent multivariate medical time series data. They looked into using the Loss Sensitive GAN (LS-GAN) objective function as a new loss function and adapted it to multivariate time series generation. To evaluate the proposed approach, the multivariate Dynamic Time Warping (MVDTW) and MMD metrcis are used.

Golany \etal  \cite{golany2020improving} developed a set of generative models based on the DCGAN architecture in order to synthesize different classes of ECG heartbeats. To assess the quality of the synthetic data, the authors used it as additional training data to evaluate its impact on the performance of neural sequence classifier.

Shaker \etal \cite{shaker2020generalization} developed a fully-connected layered GAN to generate ECG heartbeats. The synthetic data were combined with the training set to train two deep CNNs classification approaches to classify 15 arrhythmia types from the MIT-BIH dataset.

Yang \etal \cite{yang2021proegan} proposed a gradually growing generative framework called ProEGAN-MS. Beginning with a reduced resolution of ECG signals, the two networks progressively grow throughout the training process by incrementing the convolutional layer appropriate to the features extraction. The authors assessed the fidelity and diversity of the synthetic data, as well as its impact on the performance of classification algorithms.

Yi \etal \cite{XIA2023104276} introduced the TCGAN framework for ECG generation. The proposed architecture consists of a transformer generator and CNN discriminator. The generated ECG heartbeats were utilized as supplementary data alongside the real dataset to address the issue of data imbalance in ECG classification.

In \cite{QIN2023102489}, Jing \etal proposed the ECG-ADGAN model for ECG synthesis and cardiac abnormalities detection. A Bi-directional Long-Short Term Memory (Bi-LSTM) layer is used in the GAN generator and a mini-batch discrimination is used for the training of the proposed framework. The generated data were used with the real data for one-class ECG classification task.

\subsection{ECG Generation Using Customized GANs}

Golany \etal \cite{golany2019pgans} introduced a Personalized GAN (PGAN) to generate patient-specific synthetic ECG signals. A customized loss was introduced in the training process of the generator to imitate the real heartbeats morphology. This loss is a combination of the standard cross-entropy loss and the mean squared error (MSE) which aims to produce natural waveforms similar to the real waveforms detected by the NeuroKit algorithm. The generated ECG signals were combined with the real training set to train patient-specific arrhythmia classifier to demonstrate the effect of adding synthetic data on the classification performance.

Later, Golany \etal \cite{golany2020simgans} presented the SimGAN, a GAN framework that has been enhanced with supplementary knowledge from an ECG simulator. This ECG simulator was defined by three ordinary differential equations (ODE) and aims to understand the dynamic nature of ECG signal and to represent its heart dynamics. A specific loss was added to the standard cross-entropy loss of the generator optimization in order to synthesize fake heartbeats that are morphologically close to the heartbeats produced from the ECG simulator.

More recently, Golany \etal \cite{golany2021ecg} proposed the ECG-ODE-GAN framework where the generator is defined as an ODE to learn the dynamics of ECG signals. The authors also demonstrated how to incorporate physical considerations into the ECG-ODE-GAN by introducing physical parameters that characterize the ECG signals as supplementary input to the generator. To assess their model, the generated data were used to train heartbeats classifier.

In our previous work \cite{nour2021Disentangling}, we introduced a new GAN framework for generating ECG heartbeats. We demonstrated how to integrate prior knowledge about ECG waveforms by modeling the ECG signals as shape clusters (\ie signal averages) that serve as references to represent ECG patterns in the training distribution. Separating between the amplitude and temporal variations in the data modeling was also introduced in the generation process to improve the modeling of the ECG complex characteristics. The proposed approach was evaluated by combining the synthetic heartbeats as additional data with the real training set to train three arrhythmia classification baselines.

\subsection{Discussion}
Table \ref{related-work} summarizes the previously discussed approaches. Although the solutions belonging to the first category can generate ECG heartbeats, they are still limited to synthesize ``realistic" ECG signals ``with morphological proprieties". This is due to the use of the standard GAN architecture that does not include specific knowledge related to ECG signals.

% Recent solutions \cite{golany2019pgans,golany2020simgans,golany2021ecg,nour2021Disentangling} are based on GAN models enriched with prior knowledge of ECG signal patterns to generate synthetic ECG heartbeats with real morphologies. They typically rely on integrating handcrafted physics knowledge. As a result, methods for specifying data properties were required during the generation process such as using specific simulators, adding complementary algorithms to identify specific patterns in the waveforms or using handcrafted decomposition for data modeling.
% In contrast to previous solutions, we present an automated method to generate ECG signals with no feature specification. Our method is based on a 2-D statistical modeling of ECG signals.

Due to complex dynamic nature of ECGs, recent solutions \cite{golany2019pgans,golany2020simgans,golany2021ecg} are based on GAN models enriched with prior knowledge of ECG signal patterns to generate synthetic ECG heartbeats with real morphologies. They typically rely on integrating handcrafted physics knowledge. As a result, methods for specifying data properties were required during the generation process such as using specific simulators, adding complementary algorithms to identify specific patterns in the waveforms. In our previous work \cite{nour2021Disentangling}, we suggested employing ECG anchors to leverage a learned prior knowledge into the generation process. However, in this solution, we considered 1-D ECG pattern dynamics modeling by dissociating the temporal and amplitude variations of ECG signals. We acknowledge that this decomposition of variations may not accurately capture the complexity of patterns and variations observed in ECG signals.

In contrast to all previous solutions, we believe that ECG signal generation requires an advanced prior knowledge modeling of the ECG shape and dynamics for a better control over the generation process. Statistical shape models have demonstrated their effectiveness in capturing shape variations in various applications, including facial analysis. Therefore, in this work, we propose the first ECG generation approach that integrates statistical shape modeling as prior knowledge of ECG signal dynamics and shape variations. This integration offers several advantages.

Firstly, the statistical shape model provides a compact representation of the shape variations present in ECG signals. This enables the GAN architecture to generate a wide range of samples that align with the inherent shape characteristics and dynamics of real ECG signals. This integration enables the generation approach to effectively learn and produce realistic variations in shape while preserving the essential structural properties of the ECG waveform.

Moreover, the integration of a statistical shape model allows for the encoding of prior knowledge regarding shape variations. This prior knowledge serves as a valuable guidance mechanism for GANs during the generation process. By incorporating the statistical shape model, the GAN can leverage this prior knowledge to improve the fidelity and accuracy of the generated ECG signals. This ensures that the generated waveforms are not only diverse but also adhere to the expected shape characteristics of real ECG signals, enhancing the realism of the generated outputs.

\clearpage

\FloatBarrier
\begin{sidewaystable}[h]
\centering
\caption{Overview of the related work.}
\vspace*{2mm}
\label{related-work}

\resizebox{22cm}{!}{
\begin{tabular}{|l|ccccccccc|}
\hline
\multirow{2}{*}{}           &\multirow{2}{*}{Papers} & \multirow{2}{*}{Year}                                          & \multirow{2}{*}{\begin{tabular}[c]{@{}c@{}}Input \\ Generator\end{tabular}}     
& \multicolumn{2}{c}{\begin{tabular}[c]{@{}c@{}}Output\\  Generator\end{tabular}}
& \multicolumn{2}{c}{\begin{tabular}[c]{@{}c@{}}Prior Knowledge \\ modeling\end{tabular}} & \multirow{2}{*}{Evaluation}   & \multirow{2}{*}{Database}              
\\ \cline{5-6} \cline{7-8} &&  &  & \multicolumn{1}{c}{end to end} & decomposition  & \multicolumn{1}{c}{Handcrafted} & Learned &   &    \\ \hline

\multirow{13}{*}[-13.99ex]{\rotatebox{90}{\textbf{Standard GAN architecture}}}  &\cite{delaney2019synthesis}                      & 2019                  & noise                 & \multicolumn{1}{c}{\checkmark}             &          & \multicolumn{1}{c}{\xmark}          & \xmark         &\begin{tabular}[c]{@{}c@{}} Qualitative eval., \\ statistical metrics   \end{tabular}                                                                                 & MIT-BIH arrhythmia                                                        \\ \cline{2-10}
&\cite{zhu2019electrocardiogram}                       & 2019                   & noise                                                 & \multicolumn{1}{c}{\checkmark}             &                         & \multicolumn{1}{c}{\xmark}          & \xmark         & \begin{tabular}[c]{@{}c@{}}statistical metrics\end{tabular}                                                                               & MIT-BIH arrhythmia                                                       \\ \cline{2-10}
&\cite{wang2020accurate}                       & 2020                 & noise                                                 & \multicolumn{1}{c}{\checkmark}             &             & \multicolumn{1}{c}{\xmark}          & \xmark         & Qualitative eval.                                                                                & MIT-BIH arrhythmia                                                       \\ \cline{2-10}
&\cite{antczak2020generative}                       & 2020                                         & noise                                                 & \multicolumn{1}{c}{\checkmark}             &                        & \multicolumn{1}{c}{\xmark}          & \xmark         & \begin{tabular}[c]{@{}c@{}}Qualitative eval.\\ Denoiser ECG\end{tabular}                     & PTB-XL                                                                    \\ \cline{2-10}
&\cite{nankani2020investigating}                       & 2020                            & noise, class label                                                 & \multicolumn{1}{c}{\checkmark}             &                        & \multicolumn{1}{c}{\xmark}          & \xmark         & \begin{tabular}[c]{@{}c@{}} statistical \\metrics\end{tabular}                                                                              & MIT-BIH arrhythmia                                                       \\ \cline{2-10}

&\cite{hazra2020synsiggan}                       & 2020                     & noise                                                 & \multicolumn{1}{c}{\checkmark}             &                        & \multicolumn{1}{c}{\xmark}          & \xmark         & \begin{tabular}[c]{@{}c@{}}Qualitative eval., \\statistical metrics\end{tabular} & MIT-BIH arrhythmia                                                      \\ \cline{2-10}
&\cite{brophy2020synthesis}                       & 2020                 & noise                                                 & \multicolumn{1}{c}{\checkmark}             &                         & \multicolumn{1}{c}{\xmark}          & \xmark         &  \begin{tabular}[c]{@{}c@{}}Qualitative eval., \\ statistical metrics   \end{tabular}                                                                               & MIT-BIH arrhythmia                                                       \\ \cline{2-10}
&\cite{golany2020improving}                       & 2020                   & noise                                                 & \multicolumn{1}{c}{\checkmark}             &                          & \multicolumn{1}{c}{\xmark}          & \xmark         &\begin{tabular}[c]{@{}c@{}}Qualitative eval., \\ ECG classification \end{tabular}                                                                                     & MIT-BIH arrhythmia                                                       \\ \cline{2-10}
&\cite{shaker2020generalization}                       & 2020                       & noise                                                 & \multicolumn{1}{c}{\checkmark}             &                         & \multicolumn{1}{c}{\xmark}          & \xmark         & \begin{tabular}[c]{@{}c@{}}Qualitative eval., \\ ECG classification \end{tabular}                                                                                   & MIT-BIH arrhythmia                                                        \\ \cline{2-10}
&\cite{brophy2021multivariate}                      & 2021                & noise                                                 & \multicolumn{1}{c}{\checkmark}             &                         & \multicolumn{1}{c}{\xmark}          & \xmark         & \begin{tabular}[c]{@{}c@{}}Qualitative eval., \\statistical metrics \end{tabular}                                                                                     & \begin{tabular}[c]{@{}c@{}}MIT-BIH arrhythmia,\\ MIT-BIH NSR\end{tabular} \\ \cline{2-10}

&\cite{yang2021proegan}                      & 2021                     & noise                                                 & \multicolumn{1}{c}{\checkmark}             &                         & \multicolumn{1}{c}{\xmark}          & \xmark         & \begin{tabular}[c]{@{}c@{}}Qualitative eval., \\ ECG classification,\\ statistical metrics\end{tabular}             & MIT-BIH arrhythmia                                                        \\ \cline{2-10}
&\cite{XIA2023104276}                      & 2023               & noise                                                 & \multicolumn{1}{c}{\checkmark}             &                          & \multicolumn{1}{c}{\xmark}          & \xmark         &\begin{tabular}[c]{@{}c@{}}Qualitative eval., \\ ECG classification \end{tabular}                                                                                     & MIT-BIH arrhythmia                                                       \\ \cline{2-10}
&\cite{QIN2023102489}                      & 2023           & noise                                                 & \multicolumn{1}{c}{\checkmark}             &                       & \multicolumn{1}{c}{\xmark}          & \xmark         & ECG classification                                                                                    & MIT-BIH arrhythmia                                                        \\ \hline
\multirow{5}{*}[-0.9ex]{\rotatebox{90}{ \centering\textbf{Customized GAN architecture }}}&\cite{golany2019pgans}                      & 2019              & noise                                                 & \multicolumn{1}{c}{\checkmark}             &                         & \multicolumn{1}{c}{\checkmark}      &                               & ECG classification                                                                                    & MIT-BIH arrhythmia                                                        \\ \cline{2-10}
&\cite{golany2020simgans}                      & 2020                 & noise                                                 & \multicolumn{1}{c}{\checkmark}             &                         & \multicolumn{1}{c}{\checkmark}      &                               & \begin{tabular}[c]{@{}c@{}}Qualitative eval., \\ ECG classification \end{tabular}                          & MIT-BIH arrhythmia          \\ \cline{2-10}
&\cite{golany2021ecg}                      & 2021                  & \begin{tabular}[c]{@{}c@{}}Physical\\  parameters of ECG beat,\\  initial voltage \\ value of ECG beat\end{tabular} &             &      \begin{tabular}[c]{@{}c@{}}  Ordinary Differential \\ Equations   \end{tabular}             &  \multicolumn{1}{c}{\checkmark}    &                                & \begin{tabular}[c]{@{}c@{}}Qualitative eval., \\ ECG classification \end{tabular}                                                                                   & MIT-BIH arrhythmia                                                        \\ \cline{2-10}
&\cite{nour2021Disentangling}                      & 2022                & noise, class label                                    && \begin{tabular}[c]{@{}c@{}}Temporal and \\ amplitude \\ variations\end{tabular} &      &                \multicolumn{1}{c}{\checkmark}                & \begin{tabular}[c]{@{}c@{}}Qualitative eval., \\ ECG classification \end{tabular}                                                                                   & MIT-BIH arrhythmia                                                    \\ \cline{2-10}
&Our approach                    & 2023                 & noise, class label                                    & & \begin{tabular}[c]{@{}c@{}}2-D shape \\ variations\end{tabular}         & \multicolumn{1}{c}{}                               & \checkmark     & \begin{tabular}[c]{@{}c@{}}Qualitative eval., \\ Cardiologists eval., \\ ECG classification, \\ statistical metrics \end{tabular}                                                                                    & MIT-BIH arrhythmia           \\ \cline{2-10} \hline
\end{tabular}
}

\end{sidewaystable}
\FloatBarrier

\clearpage
\begin{figure*}[th]
\centering
\includegraphics[width=\textwidth]{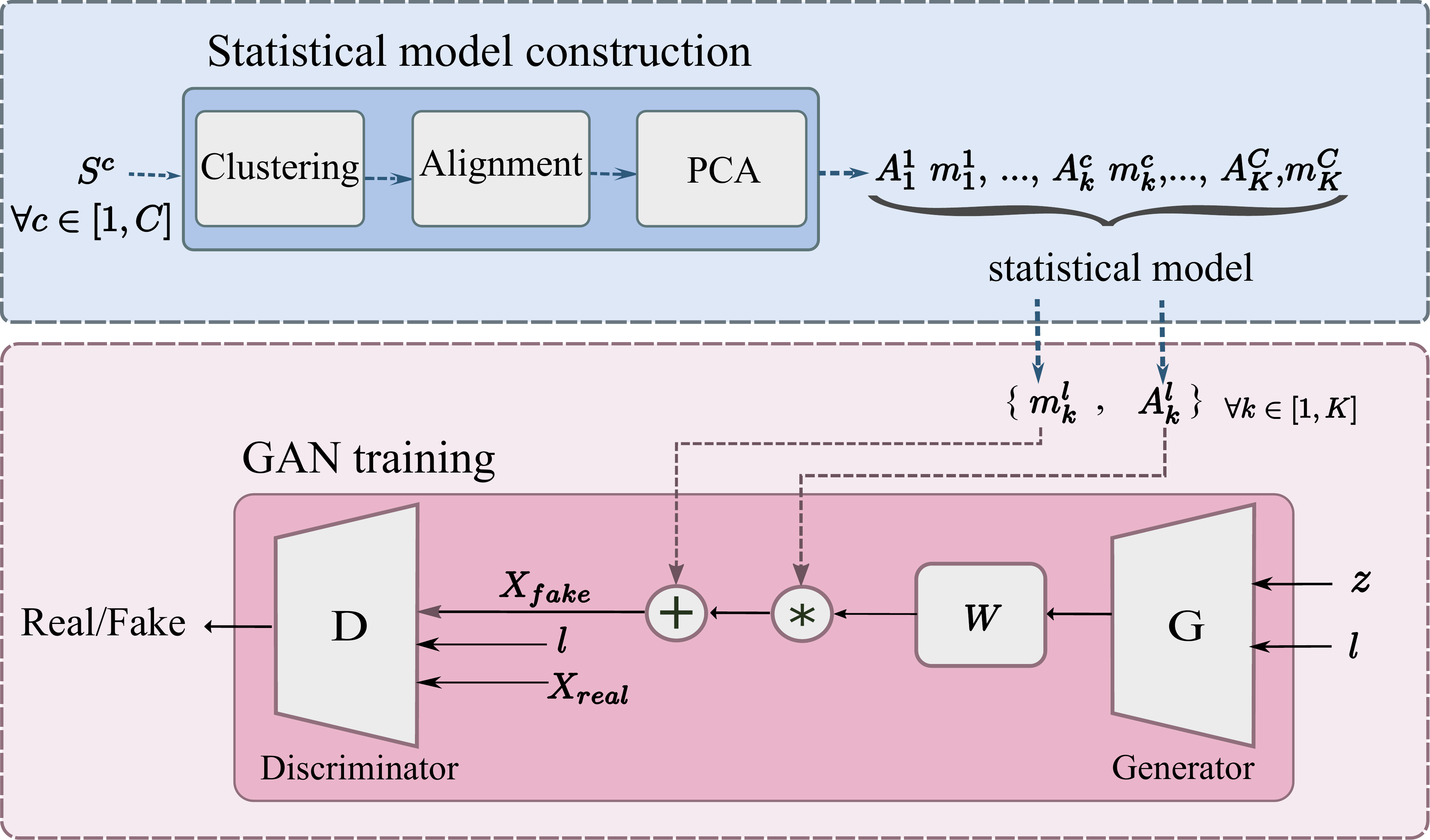}
\centering\caption{Workflow of the proposed approach.}
\label{pipeline}
\end{figure*}

% =============================================
% # III. Proposed Model #
% =============================================

\section{Proposed Approach\label{sec:proposed_approach}}

The workflow of our proposed method is shown in Figure \ref{pipeline}. We first start with a processing block designed for statistical modeling. Then, we move on to the training phase of our GAN model. The generator takes a noise vector and data labels as input to generate a specific linear combination that will be used to the output of the statistical model. Afterwards, the real ECG heartbeats, as well as the synthetic heartbeats, are passed to the discriminator, which attempts to distinguish between the real and fake heartbeats. The underlying principles are detailed below.

\subsection{Statistical Model Construction}

The aim of our statistical model is to accurately model ECG signal shape variations. It consists of three consecutive steps: clustering, alignment, and statistical modeling for signal decomposition. The pseudo-code of this phase is outlined in Algorithm \ref{pca2}. First, the training dataset is composed of $C$ subsets of signals  $\{\Sb\}^c$ grouped by class, where $c \in [1,C]$ is the class index and $\Sb$ refers to one signal.
As discussed above, the ECG signal is represented by its 2-D point coordinates (\ie position in time and amplitude, respectively) where $\Sb \in \mathbb{R}^{2\times T}$ and $T$ is the length of the signal.

The signals of each subset are clustered into fixed $K$ clusters of signals noted by $\{\Sb\}^c_k$, where $k$ is the cluster index. Without loss of generality, we use the K-Means clustering method \cite{kmeans}.

% \begin{algorithm}
% \SetAlgoLined
% \caption{Statistical model construction.}

% \label{pca2}
% \SetAlgoLined
% %\KwInput{Base}
% \KwData{Number of classes $C$, Number of clusters $K$}
% \For{c:=1 to C} {   
%             %$\{\Sb\}^c$= selectSamples($\{\Sb\}$,c)\; 
%             $\{\Sb\}^c_k$ = K-Means($\{\Sb\}^c$,$K$)\;
%             \For{k:=1 to K} {   
%                         $\{\pb\}^c_k$ = DTAN($\{\Sb\}^c_k$)\;
%                         $\{\hat{\Sb}\}^c_k$ = interpolate($\{\Sb\}^c_k$, $\{\pb\}^c_k$)\;
%                         % $\Xb^c_k$ = $\begin{bmatrix} \{\hat{\Sb}[0]\}^c_k & \{\hat{\Sb}[1]\}^c_k  \end{bmatrix}$\;
%                         $\Xb^c_k$ = $\left[\{\hat{\Sb}[0]\}^c_k | \{\hat{\Sb}[1]\}^c_k  \right]$\; \label{lst:line:X}
%                         $\mb^c_k$,$\Ab^c_k$ = PCA($\Xb^c_k$) \;
%             }
% }
% %
% \end{algorithm}

\begin{figure}[tbp]
% \begin{figure}[th]
\centering
\includegraphics[width=0.5\textwidth]{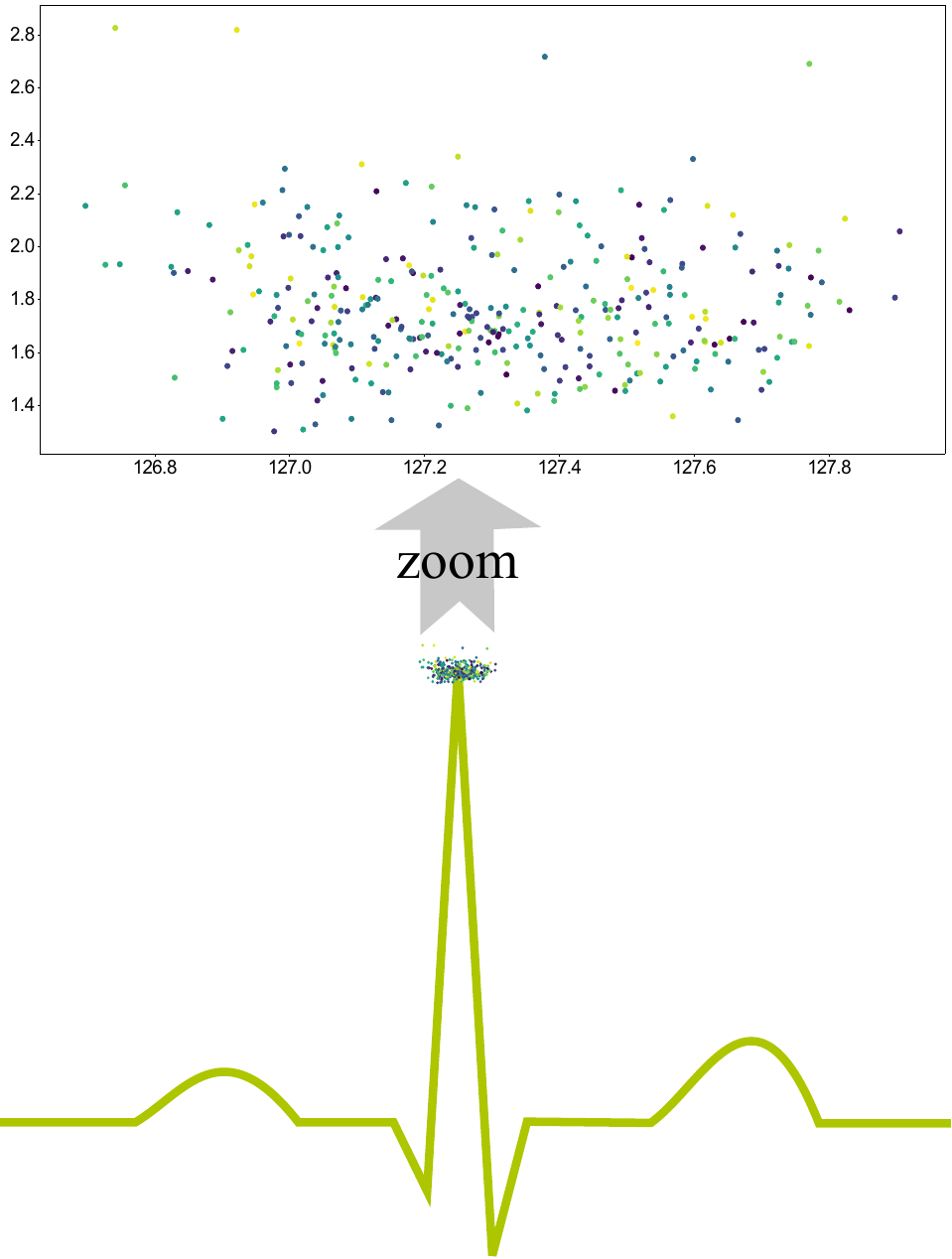}
\caption{A set of homologous points corresponding to the peak R localized in different signals that belong to the same cluster.}
\label{2D_variations}
\end{figure}

\begin{figure}[tbp]
\centering
\includegraphics[width=\textwidth,height=0.35\textheight ]{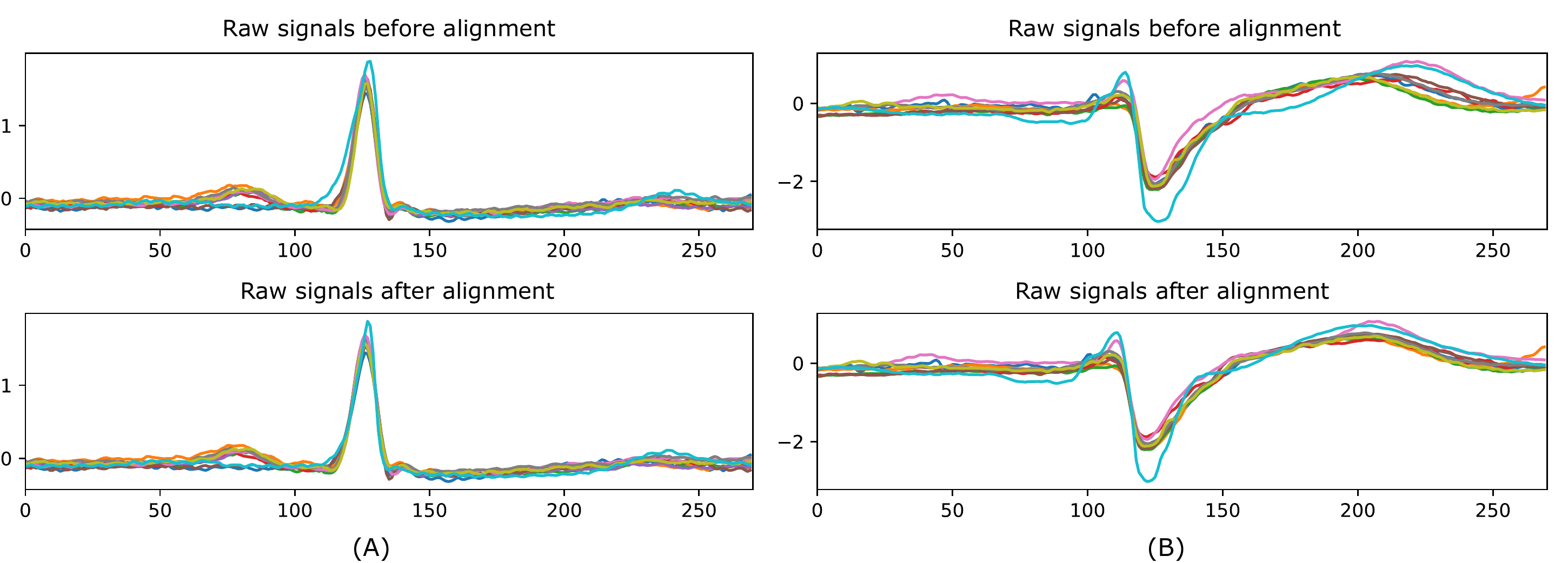}
\caption{(A): Examples of signals before and after alignment taken from the training dataset of the normal class (N).
(B): Examples of signals before and after alignment taken from the training dataset of pathological class.}
\label{alignment1}
\end{figure}

To model the time and amplitude variations of homologous points inside a cluster of signals, we must first localize the position of each homologous point in every signal of $\{\Sb\}^c_k$. As an example, Figure \ref{2D_variations} shows a plot of homologous points corresponding to the peak R localized in different signals that belong to a same cluster.
\begin{algorithm}
\SetAlgoLined
\caption{Statistical model construction.}

\label{pca2}
\SetAlgoLined
%\KwInput{Base}
\KwData{Number of classes $C$, Number of clusters $K$}
\For{c:=1 to C} {   
            %$\{\Sb\}^c$= selectSamples($\{\Sb\}$,c)\; 
            $\{\Sb\}^c_k$ = K-Means($\{\Sb\}^c$,$K$)\;
            \For{k:=1 to K} {   
                        $\{\pb\}^c_k$ = DTAN($\{\Sb\}^c_k$)\;
                        $\{\hat{\Sb}\}^c_k$ = interpolate($\{\Sb\}^c_k$, $\{\pb\}^c_k$)\;
                        % $\Xb^c_k$ = $\begin{bmatrix} \{\hat{\Sb}[0]\}^c_k & \{\hat{\Sb}[1]\}^c_k  \end{bmatrix}$\;
                        $\Xb^c_k$ = $\left[\{\hat{\Sb}[0]\}^c_k | \{\hat{\Sb}[1]\}^c_k  \right]$\; \label{lst:line:X}
                        $\mb^c_k$,$\Ab^c_k$ = PCA($\Xb^c_k$) \;
            }
}
\end{algorithm}
We use a Diffeomorphic Temporal Alignment Net (DTAN) \cite{weber2019diffeomorphic} to align the signals $\{\Sb\}^c_k$.
A DTAN is an efficient way to align time series via flexible temporal transformer layers. Compared to other existing solutions for data alignment, a DTAN shows better results on different types of time series data including ECG signals of the ECGFiveDays dataset \cite{UCRArchive}. 
Figure \ref{alignment1}(a) and Figure \ref{alignment1}(b) show examples of normal and pathological ECG signals before and after alignment.

In addition to the aligned signals, DTAN provides the positions $\{\pb\}^c_k$ that will be used later to interpolate their corresponding time and amplitude values from $\{\Sb\}^c_k$ yielding to $\{\hat{\Sb}\}^c_k$. Then, the set of signals $\{\hat{\Sb}\}^c_k$ is arranged in a matrix $\Xb^c_k \in \mathbb{R}^{N^c_k\times 2T}$ by concatenating their time and amplitude values.
%, as described online~\ref{lst:line:X}. 
$N^c_k$ is the total number of signals in $\{\Sb\}^c_k$.

The final step is to apply Principal Component Analysis (PCA) to decompose $\Xb^c_k$ into a mean vector $\mb^c_k \in \mathbb{R}^{2T}$ (\ie shape average of signals) and variation matrix $\Ab^c_k \in \mathbb{R}^{B^c_k\times 2T}$ regrouping $B^c_k$ eigenvectors. This decomposition allows us to represent any signal belonging to a cluster $k$ and a class $c$ within a linear combination between $\mb^c_k$ and $\Ab^c_k$.

\subsection{Generative Adversarial Network Architecture}

The generator $G(\zb,l)$ receives as input a random noise vector $\zb$ sampled from a normal distribution $\Nc$(0,1) and a label $l \in [1,C]$ conditioning the class of the signals to generate, and outputs a matrix $\Wb$. $\Wb$ gathers the weights used for a linear combination of the eigenvectors in $\Ab^c_k$ where $c=l$. It is worth noticing that $\Ab^c_k$ matrices have variable number of eigenvectors $B^c_k$. For the sake of readability, we define $\Wb \in \mathbb{R}^{K\times \underset{ c,k }{\max}{(B^c_k)}}$ and we add row zero-padding on $\Ab^c_k$ matrices so that $\Ab^c_k$ becomes $\in \mathbb{R}^{\underset{ c,k }{\max}{(B^c_k)} \times 2T} \forall c \in [1,C]$ and $k \in [1,K]$. In this way, the generation of fake signals $\Xb_{fake}$ is:

\begin{eqnarray}
\label{eq:fake_sign_generation1}
\Xb_{fake} &=& \Mb_l + \Wb  \Ac_l \text{,} \\  \text{where    } 
\nonumber \Wb &=& G(\zb,l) \text{.}
\end{eqnarray}

In Equation (\ref{eq:fake_sign_generation1}), the matrix $\Xb_{fake}$ $\in \mathbb{R}^{K \times 2T}$ arranges one fake signal by each cluster. $\Mb_l$ $\in \mathbb{R}^{K \times 2T}$ is a matrix that regroups the means  $\mb^l_k, \forall~k \in [1,K]$. $\Ac_l$ $\in \mathbb{R}^{ K \times \underset{ c,k }{\max}{(B^c_k)} \times 2T}$ is a tensor gathering all $\Ab^l_k$ matrices $\forall~k \in [1,K]$.

The architecture of our generator and discriminator networks are described in Table \ref{tab:architecture_G_D}. All used convolution layers apply one-dimension (1-D) convolution. The generator's input noise vector and labels are first concatenated and then passed through three convolution layers, each followed by a batch normalization layer and a ReLU activation function. The output of the final convolution layer is then fed into a batch normalization layer, which is then processed by an FC layer to obtain the desired output shape of \Wb.
The discriminator consists of three convolution layers. A batch normalization followed by a ReLU activation function are used between these three layers. The output of the last convolutional layer is passed to a batch normalization layer and a FC layer. This is followed by an LSTM layer, three FC layers and finally a sigmoid activation function to produce the classification scores. Despite the simplicity of the proposed architecture, various architectures, including the transformer network, were investigated for both the generator and discriminator. However, no notable difference in results was observed across these architectures.

{The training of GAN is a min-max game between its components, which can be expressed as follows:
\begin{equation}
\label{ganloss}\min _ { G } \max_{ D } \underset { \zb \sim P _ { g } } { \Eb  } [ \log ( 1 - D ( G ( \zb,l ) ) ) ] + \\  \underset {  _{ \Xb_{real} \sim P_{real} } } { \Eb } [ \log ( D ( \Xb_{real},l ) ) ] 
\end{equation}

\noindent where $\Xb_{real}$ represents the input data sampled from the real distribution $P_{real}$, $D(\Xb_{real},l)$ and $D(G(\zb,l))$ are the probabilities estimated by the discriminator for classifying real instances as real and fake instances as real, respectively.
The generator attempts to minimize this loss by synthesizing fake instances that are similar to real instances to deceive the discriminator. On the other hand, the discriminator aims to maximize this loss function by maximizing both terms of this function. Therefore, it attempts to estimate a probability equal to 1 when fed $\Xb_{real}$ and 0 when fed $\Xb_{fake}$. To stabilize the training of our models, we used a Wasserstein GAN with gradient penalty loss \cite{gulrajani2017improved} instead of the original GAN loss. It consists of adding an extra penalty term to Equation (\ref{ganloss}). 
%\FloatBarrier

%\FloatBarrierThe final loss $\Lc$ is:
%
\begin{align}
\label{eq:wganloss} \nonumber \Lc= \underset {  _{ \Xb_{fake} \sim P_{fake} } } { \Eb } [D ( \Xb_{fake},l )  ] - \underset { \Xb_{real} \sim P _ { real } } { \Eb  } [ D ( \Xb_{real},l ) ] \\ 
+ \lambda \underset {  _{ \hat{\Xb} \sim P_{\hat{\Xc}} } } { \Eb } [(\|\nabla_{\hat{\Xb}} D(\hat{\Xb},l)\|_{2} -1)^2]
\end{align}

\noindent where $ \underset {  _{ \hat{\Xb} \sim P_{\hat{\Xb}} } } { \Eb } [(\|\nabla_{\hat{\Xb}} D(\hat{\Xb},l)\|_{2} -1)^2]$ is the gradient penalty, and $\lambda$ is the penalty coefficient used to weight the gradient penalty term. $P_{\hat{\Xb}}$ is the distribution obtained by randomly interpolating between samples from the distributions $P_{real}$ and $P_{fake}$.

\begin{table}[bh!]
\centering
\caption{Details of the generator and discriminator networks.}
\label{tab:architecture_G_D}
\begin{tabular}{|l|l|}
\toprule
\multirow{5}{*}[-3.5ex]{\rotatebox{90}{ \centering\textbf{Generator} }}&
Input: (\zb,$l$) \\  \cline{2-2} 
                   & \begin{tabular}[c]{@{}l@{}}Layer 1: conv1 (kernel\_size=1), BatchNorm, ReLU \\ output  (batch\_size, 16, 1) \end{tabular} \\ \cline{2-2} 
                   & \begin{tabular}[c]{@{}l@{}}Layer 2: conv1 (kernel\_size=1), BatchNorm, ReLU, \\ output (batch\_size, 32, 1)  \end{tabular} \\ \cline{2-2} 
                   & Layer 3: conv1 (kernel\_size=1), BatchNorm , output (batch\_size, 64, 1) \\ \cline{2-2} 
                   & Layer 4: FC, output (batch\_size, $K\times \underset{ c,k }{\max}{(B^c_k)})$ \\ \toprule \bottomrule 
\multirow{9}{*}[-4.5ex]{\rotatebox{90}{ \centering\textbf{Discriminator}}} &
Input: ($\Xb_{real}$/$\Xb_{fake}$, $l$) \\ \cline{2-2} 
                   & \begin{tabular}[c]{@{}l@{}}Layer 1: conv1 (kernel\_size=1), BatchNorm, ReLU \\ output  (batch\_size, 16, 1)\end{tabular} \\ \cline{2-2} 
                   & \begin{tabular}[c]{@{}l@{}}Layer 2: conv1 (kernel\_size=1), BatchNorm, ReLU \\ output (batch\_size, 32, 1) \end{tabular} \\ \cline{2-2} 
                   & \begin{tabular}[c]{@{}l@{}}Layer 3: conv1(kernel\_size=1), BatchNorm, output (batch\_size, 64, 1) \end{tabular} \\ \cline{2-2} 
                    
                   & Layer 4: FC, output (batch\_size,128) \\ \cline{2-2} 
                   & \begin{tabular}[c]{@{}l@{}}Layer 5: LSTM( hidden\_size=256, number\_layers=3) \\
    output (batch\_size,256)    \end{tabular} \\ \cline{2-2} 
                   & Layer 6: FC, Tanh, output (batch\_size,128) \\ \cline{2-2} 
                   & Layer 7: FC, Tanh, output (batch\_size,64) \\ \cline{2-2} 
                   & Layer 8: FC, Sigmoid, output (batch\_size,1) \\ \bottomrule
\end{tabular}
\end{table}

\section{Experimental Results \label{sec:experiments_and_results}}

We conducted two types of experiments to evaluate our proposed method: qualitative and quantitative evaluations. The qualitative evaluation follows two steps. First, we visually compare the real ECG heartbeats to synthetic ECG heartbeats. Then, a set of shuffled synthetic and real heartbeats are blindly evaluated by three cardiologists. The quantitative evaluation consists of assessing the effect of adding our synthetic ECG signals to the real training set on three arrhythmia classifiers performance. The synthetic ECG signals were then assessed using various metrics.
\subsection{Evaluation Database}

ECG signals taken from the MIT-BIH arrhythmia database \footnote{\href{https://physionet.org/content/mitdb/1.0.0/}{https://physionet.org/content/mitdb/1.0.0/}} were considered for our models training \cite{mitbih}. The ECG represents the main diagnostic procedure for detecting cardiac anomalies. Various cardiac abnormalities such as arrhythmia can be the sources of changes in the normal ECG patterns. The arrhythmia is a disease specified by a disorder in the cardiac rhythm accompanied by modifications of the ECG.

%The MIT-BIH arrhythmia database is the most popular dataset in arrhythmia analysis. 
The MIT-BIH arrhythmia database is widely recognized as the benchmark dataset for arrhythmia analysis, and it has been extensively used in previous studies for evaluation purposes. The database contains 48 half-hour ECG recordings, obtained from 47 patients examined between 1975 and 1979 by the BIH Arrhythmia Laboratory. Each record contains two 30-minutes ECG lead signals that have been annotated by cardiologists and digitized at 360 samples per second. Cardiac cycles can be divided into different categories that can represent normal heartbeats or arrhythmias. The P, Q, R, S, T wave patterns vary by the arrhythmia category. In this paper, three types of heartbeats were considered. The first type is the normal beats (class N). The second is the premature ventricular contraction beats (class V), and the last one is the fusion beats (class F).

\subsection{Training Settings}
Our models were implemented using the pyTorch framework and trained on an Ubuntu server version 20.04, with a GeForce GTX 1080 ti GPU having 11 GB memory. The optimization of our GAN is performed using the ADAM algorithm and a learning rate of 0.00001. The batch size used is equal to 64. The penalty coefficient $\lambda$ used in loss equation (\ref{eq:wganloss}) is set to 10 as in \cite{gulrajani2017improved,antczak2020generative}. The input $z$ of the generator is a noise vector of length 100 from the normal distribution $\Nc$(0,1 ) as in \cite{golany2021ecg,nankani2020investigating,golany2020improving,golany2019pgans}. Each ECG signal is divided into heartbeats with 270 voltage values ($T$ = 270). The $T$ values correspond to 350 ms before the R-peak and 400 ms after the R-peak. In this study, we randomly selected 70\% of the data as training dataset, while the remaining 30\% of the data was allocated for testing purposes. A low-pass filter was applied to the ECG signals to remove the noise before being fed to our models.

%%%%%%%%%%%%%%%%%%%%%%%%%%%%%%%%
\subsection{Results}
\subsubsection{Qualitative Evaluation}
%
%
% \begin{figure}[t]
% \centering

% \includegraphics[width=\textwidth]{class_N_data.eps}

% \caption{Examples of synthetic heartbeats for class N. The blue background represents a portion of the distribution of the real dataset from class N, while the green depicts synthetic heartbeats. 
% }
% \label{real_fake_n}
% \end{figure}
% %
% \begin{figure}[t]
% \centering
% \includegraphics[width=1\textwidth]{class_V_data.eps}
% \caption{Examples of synthetic heartbeats for class V. The blue background represents a portion of the distribution of the real dataset from class V, while the green depicts synthetic heartbeats.}
% \label{real_fake_ecg_v}
% \end{figure}
% \begin{figure}[h!]
% \centering
% \includegraphics[width=1\textwidth]{class_F_data.eps}
% \caption{Examples of synthetic heartbeats for class F. The blue background represents a portion of the distribution of the real dataset from class F, while the green depicts synthetic heartbeats.}
% \label{real_fake_ecg_f}
% \end{figure}

%
For the qualitative assessment of our method, a set of synthetic heartbeats were randomly selected and compared to the distributions of real heartbeats taken from the training dataset. Figure~\ref{fig:fake_real_2} shows examples of synthetic heartbeats belonging to the normal class and the pathological classes (V and F),
in addition to different clusters of the real distribution of these classes. We can observe that the synthetic heartbeats have realistic morphology and are closely similar to real distributions. In addition, the cardiac cycles demonstrate faithful dynamics according to their corresponding classes, visually proving the ability of our approach in generating realistic ECG heartbeats. 

\clearpage
\FloatBarrier
\begin{sidewaysfigure}[]
\centering
\begin{tabular}{cccccc}
& cluster 1 & cluster 2  & cluster  3 &  cluster 4 & cluster 5  \\
\multirow{3}{*}[15.9ex]{\rotatebox{90}{Class N}} & 
\includegraphics[width=0.19\textwidth]{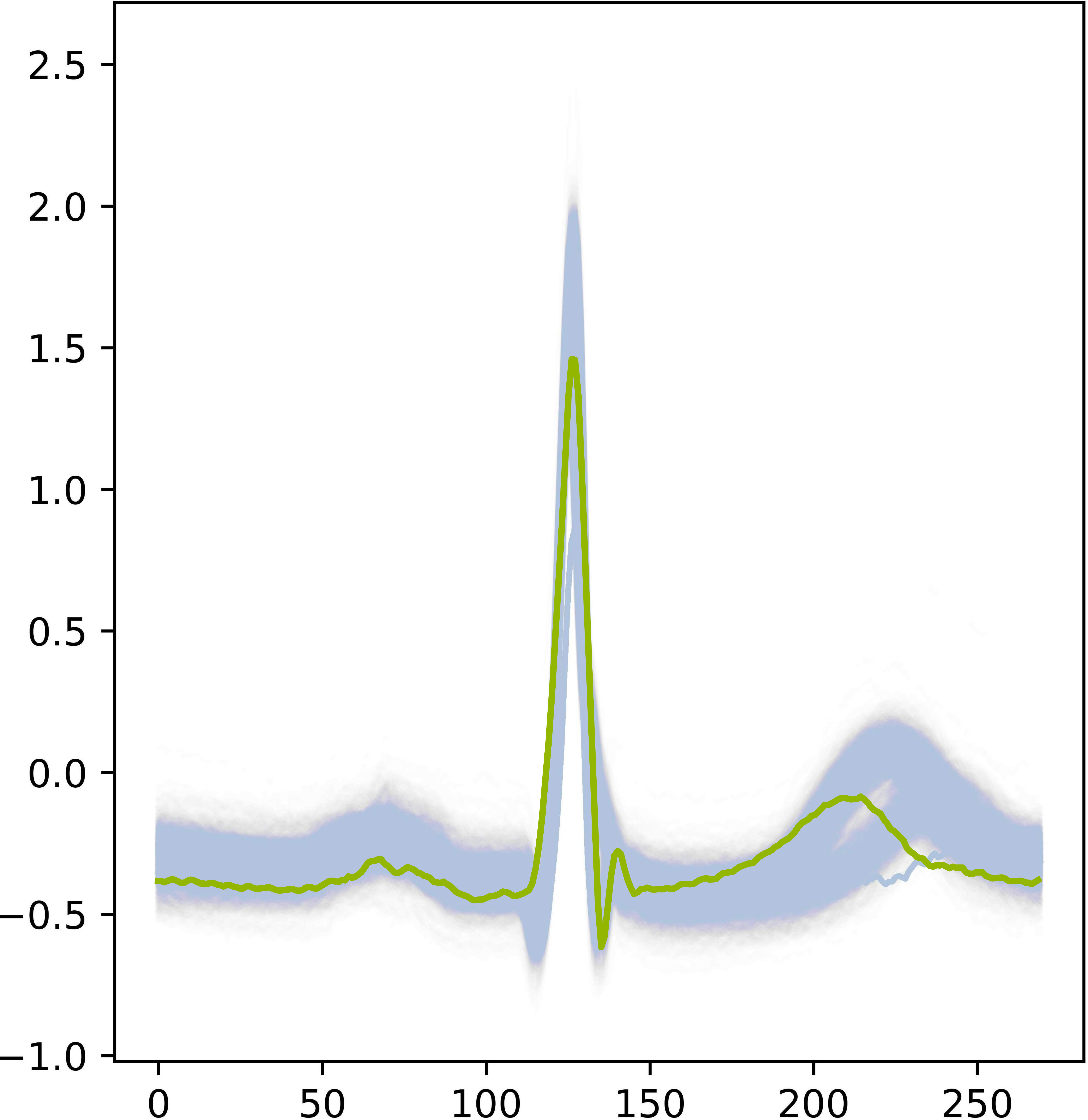} &
\includegraphics[width=0.19\textwidth]{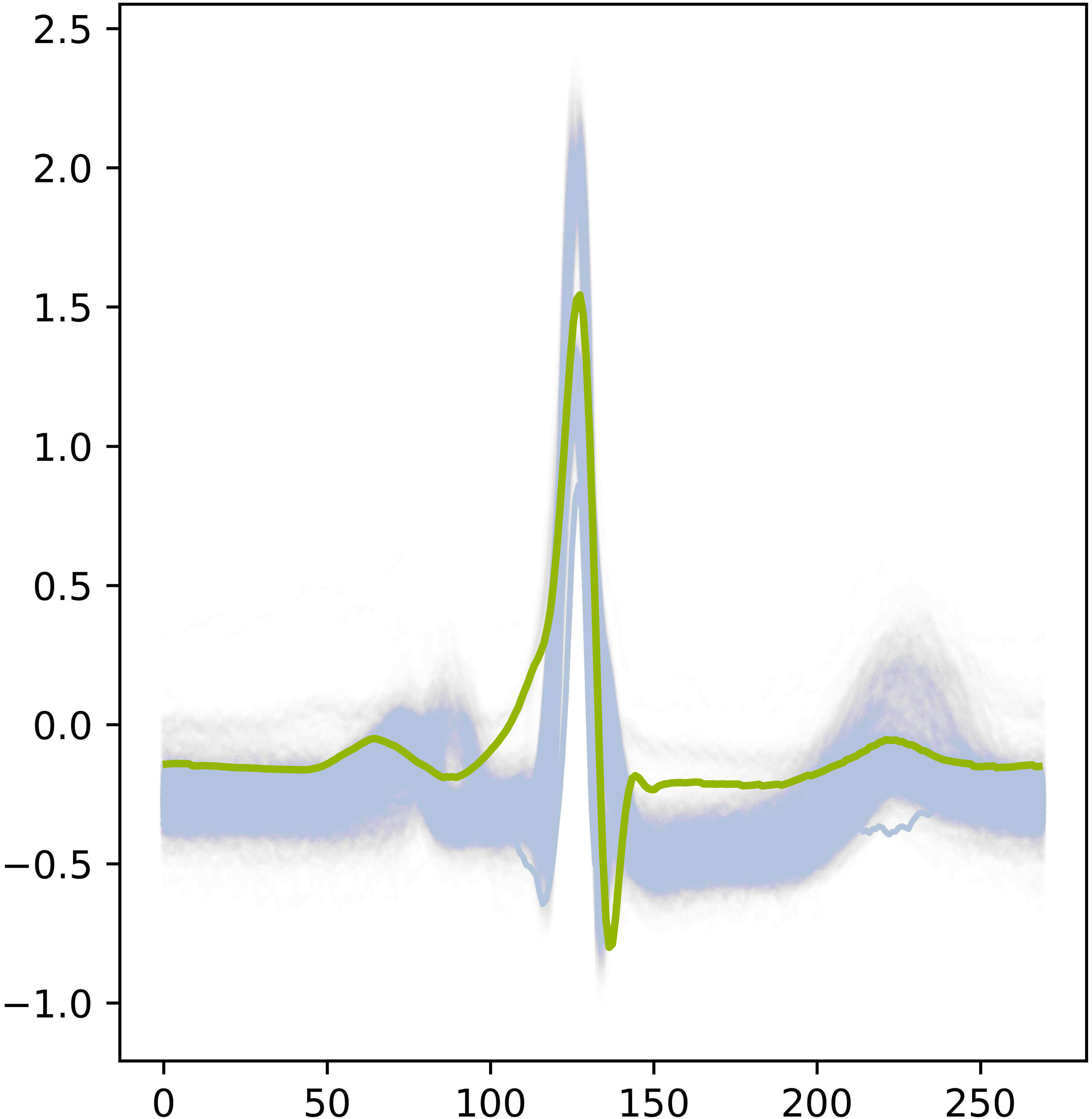} & 
\includegraphics[width=0.19\textwidth]{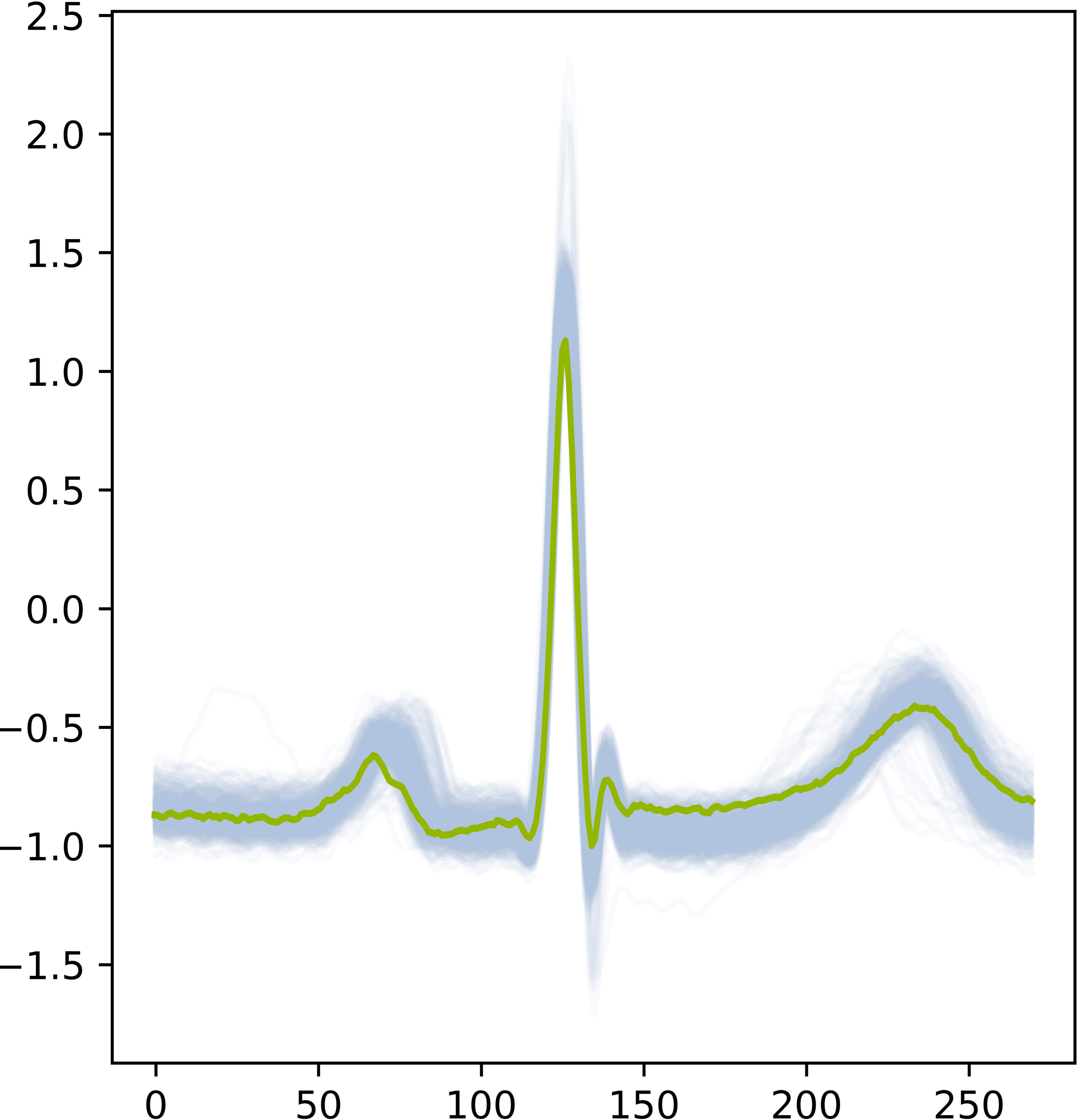} &
\includegraphics[width=0.19\textwidth]{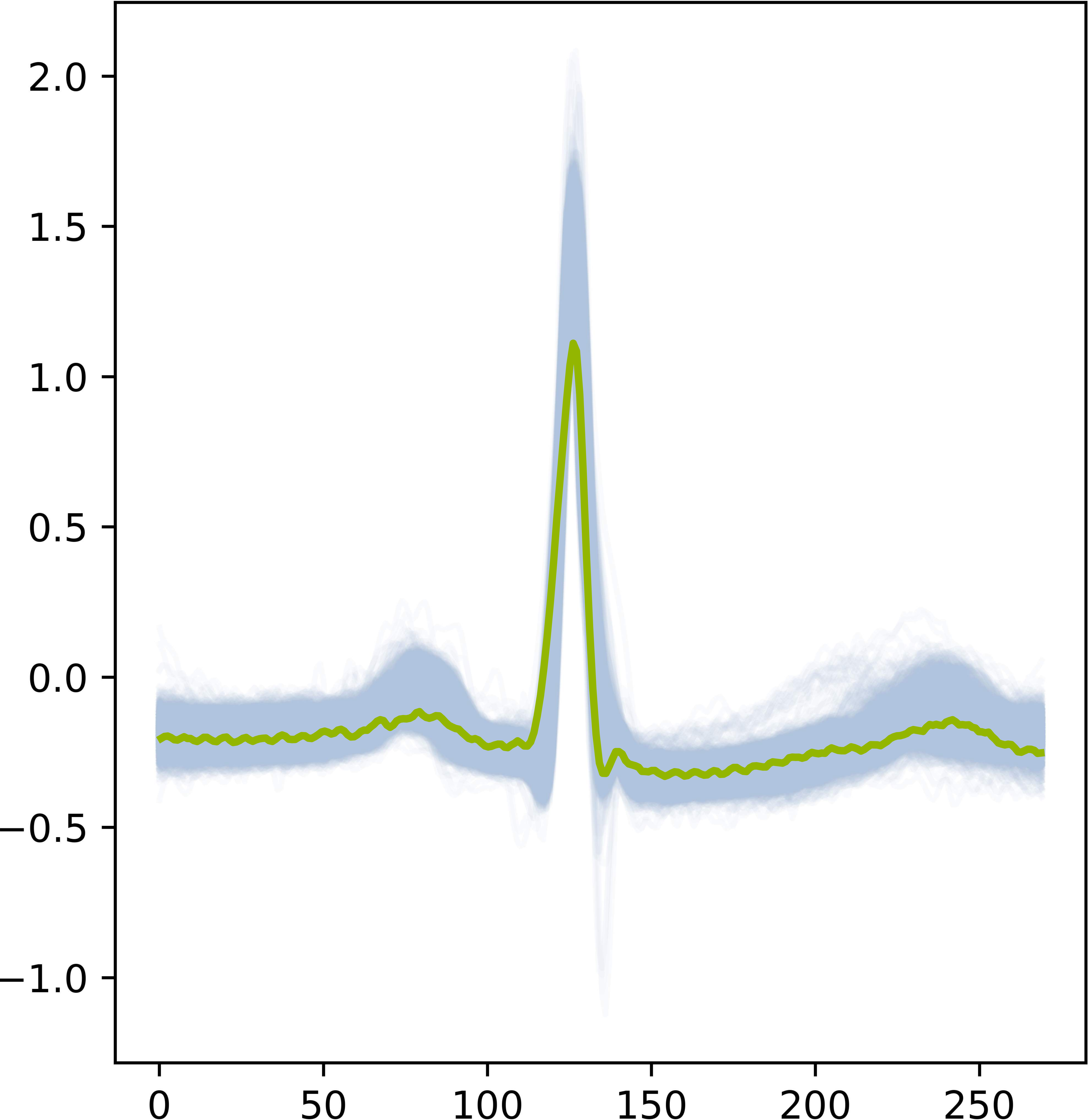} &
\includegraphics[width=0.186\textwidth]{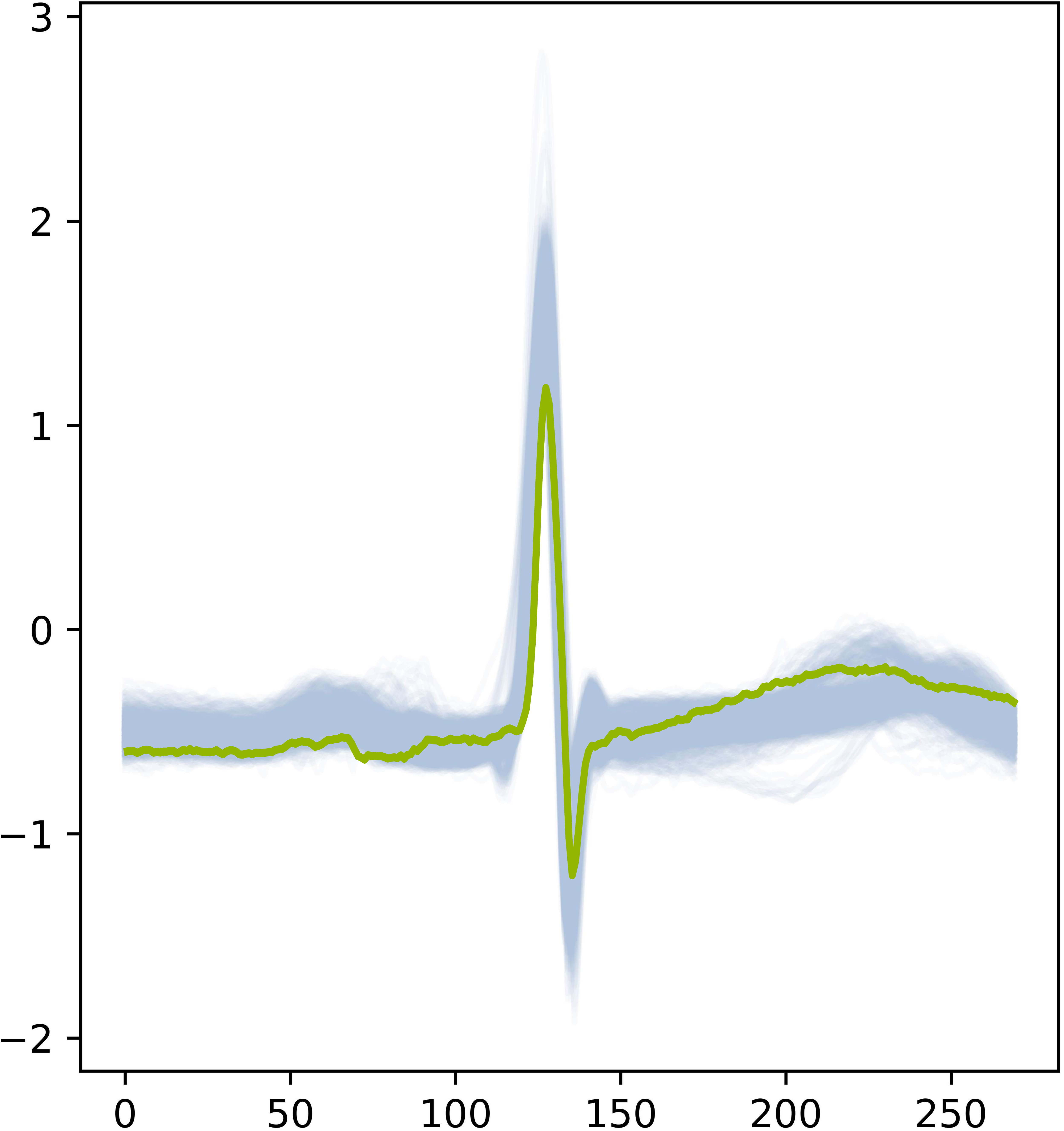} \\
\multirow{3}{*}[15.9ex]{\rotatebox{90}{Class V}} & 
\includegraphics[width=0.19\textwidth]{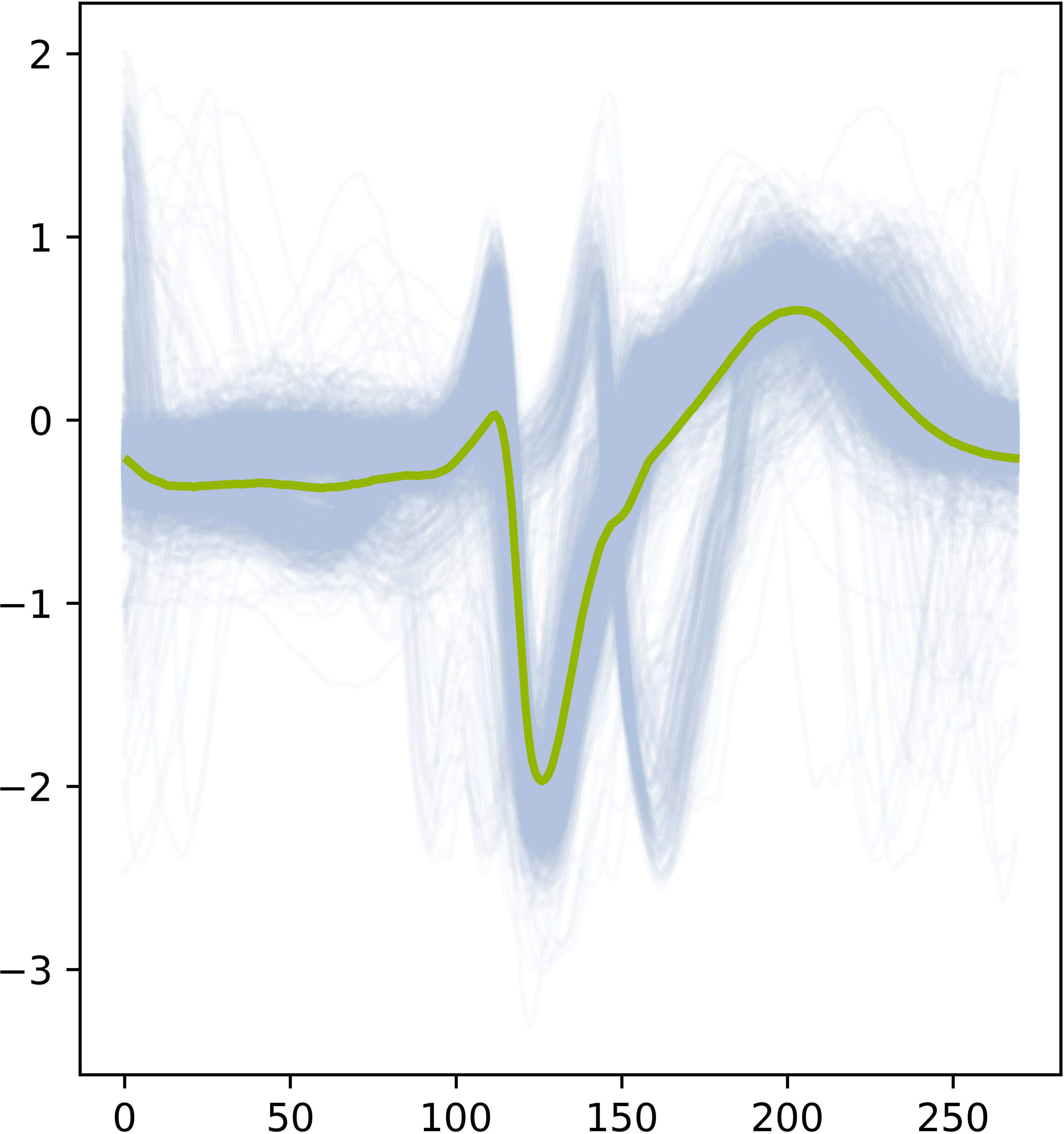} & 
\includegraphics[width=0.19\textwidth]{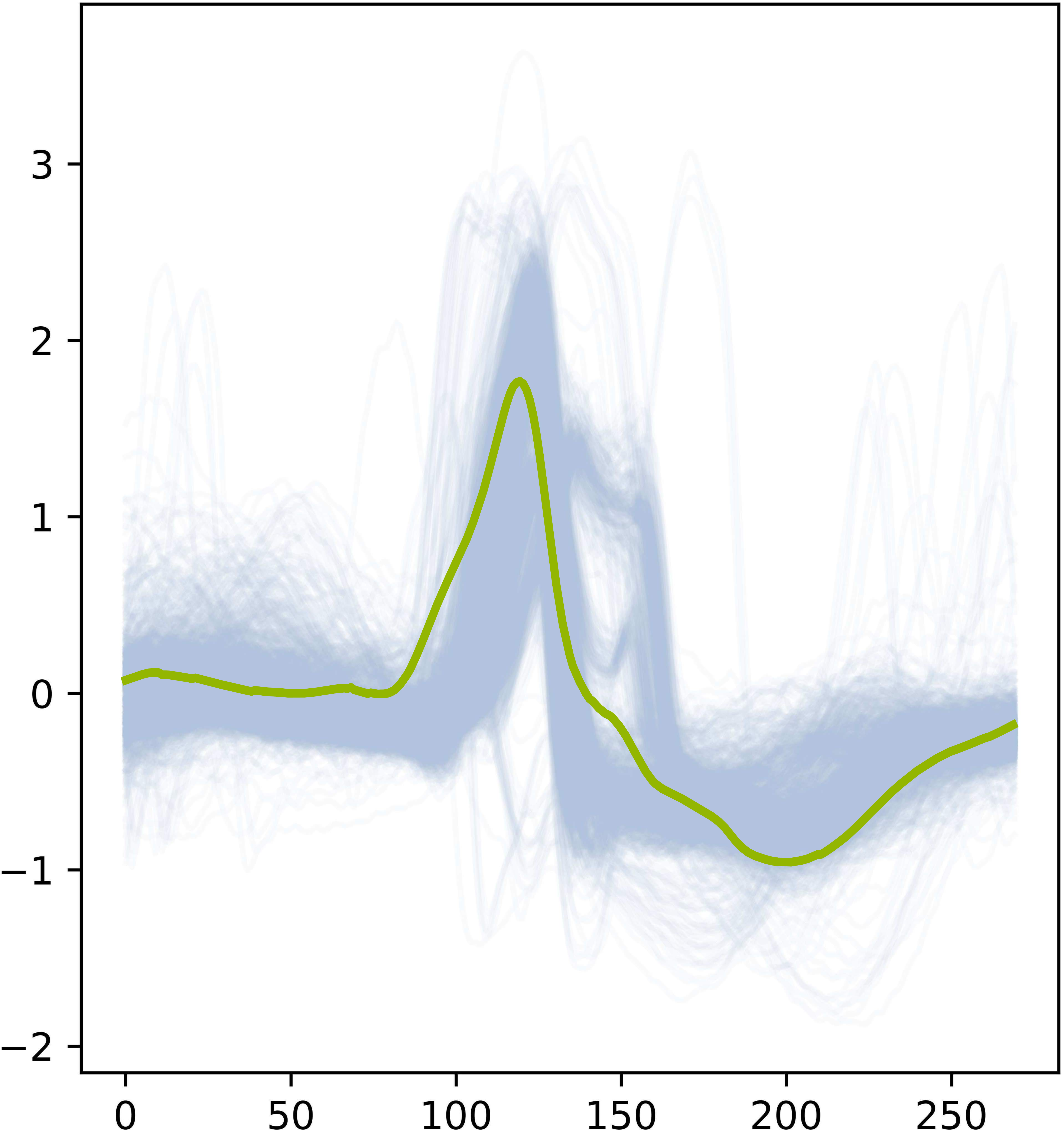} & 
\includegraphics[width=0.19\textwidth]{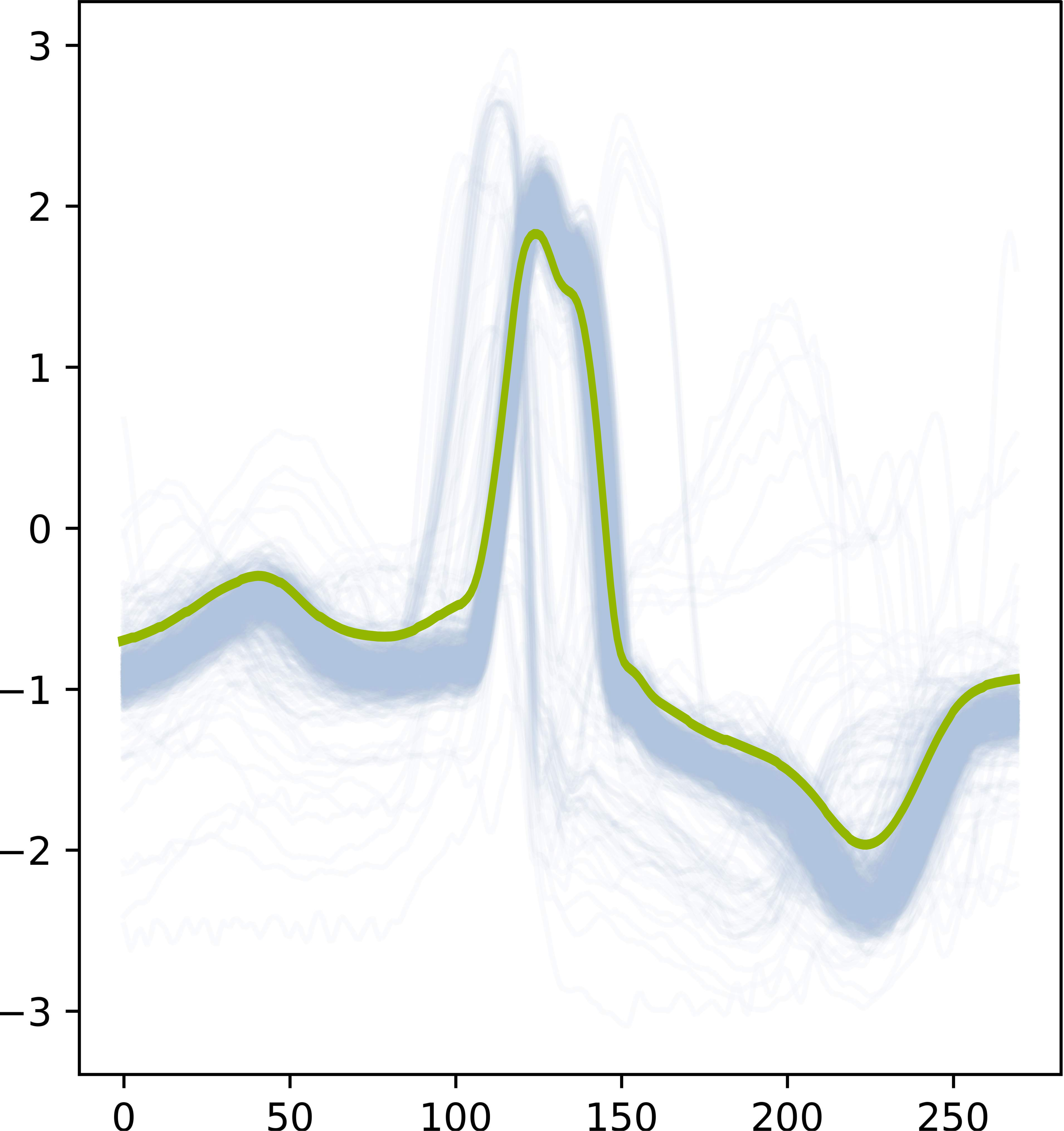} &
\includegraphics[width=0.19\textwidth]{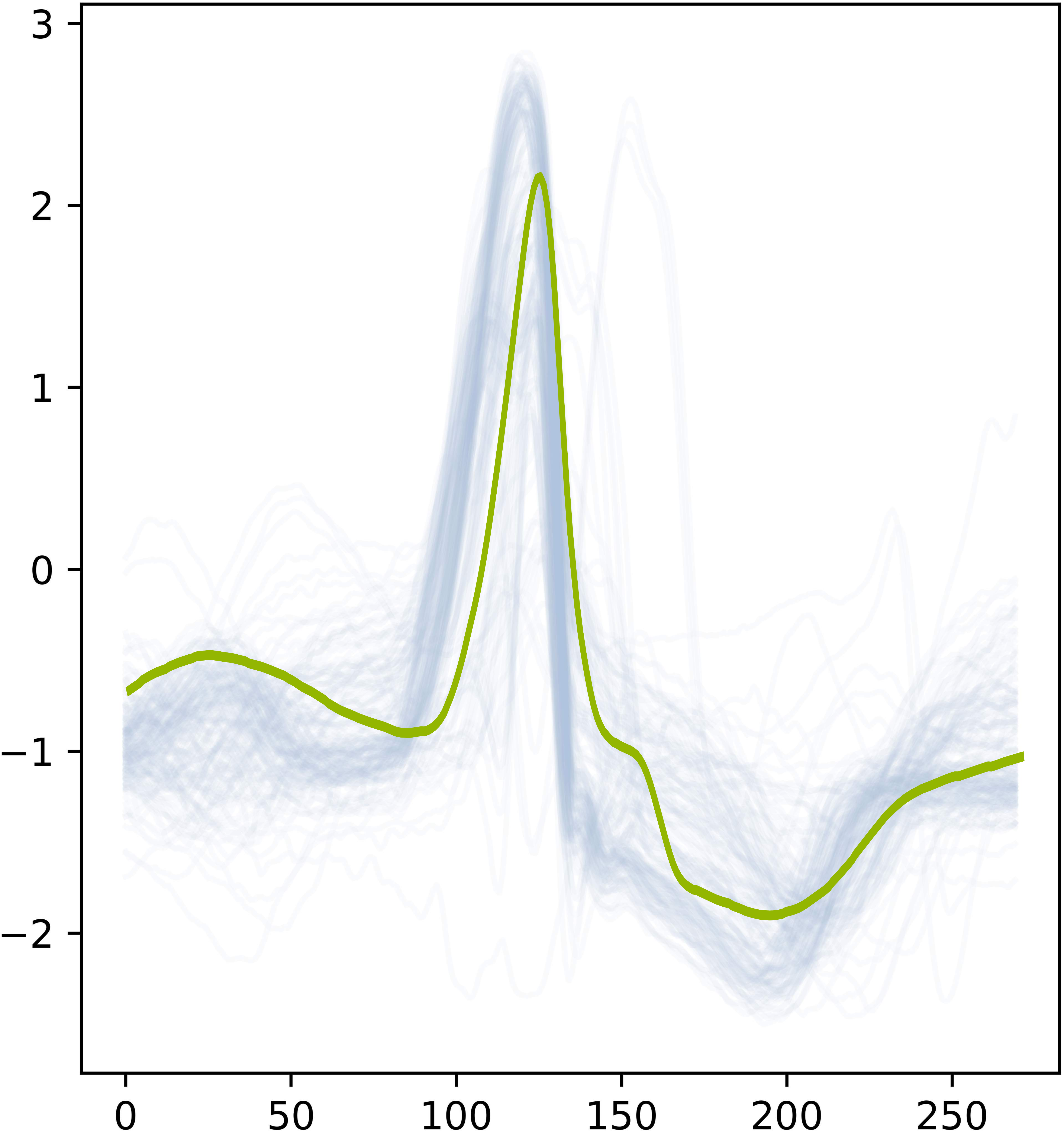} & 
\includegraphics[width=0.19\textwidth]{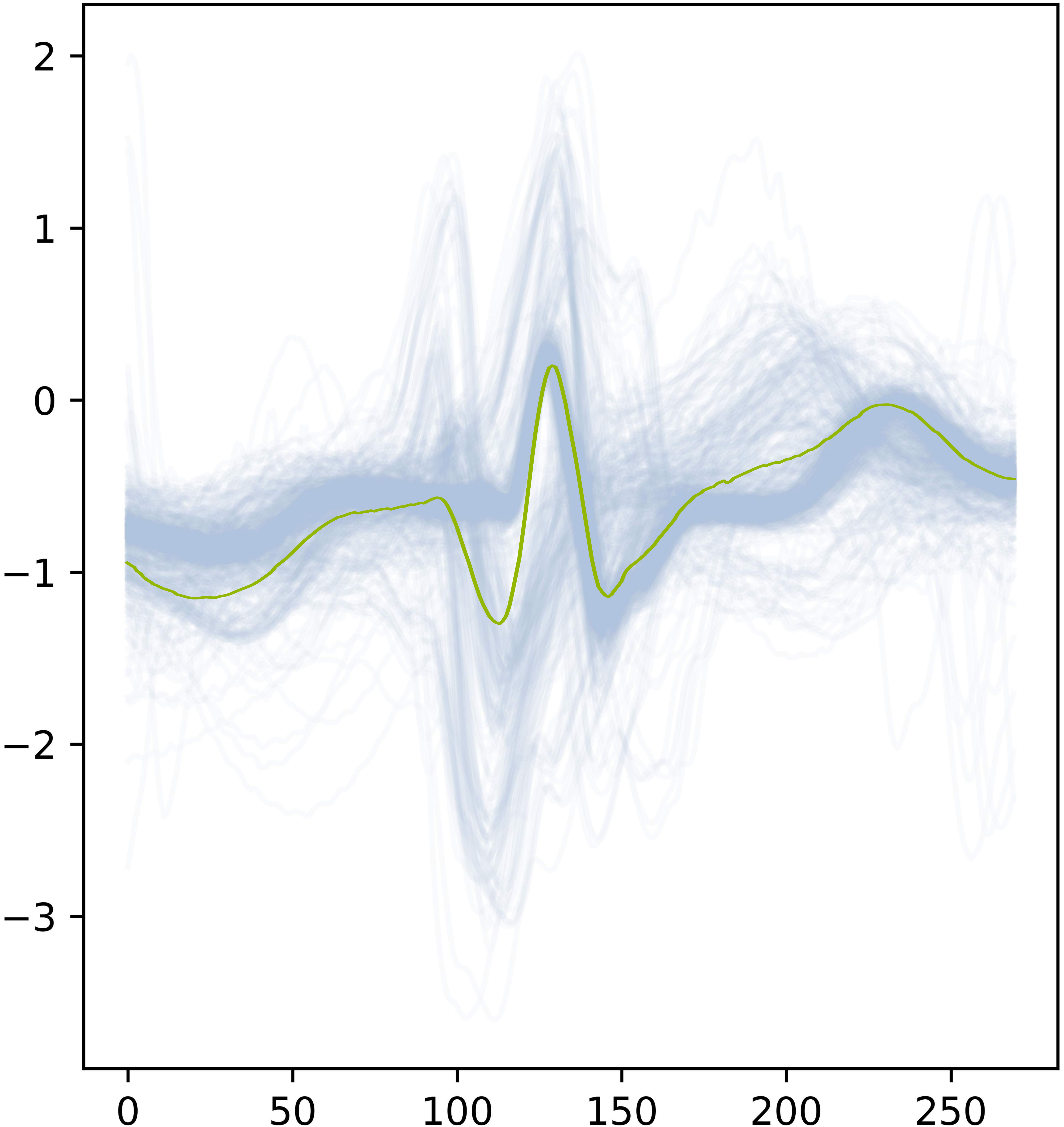} \\
\multirow{3}{*}[15.9ex]{\rotatebox{90}{Class F}} &
\includegraphics[width=0.186\textwidth]{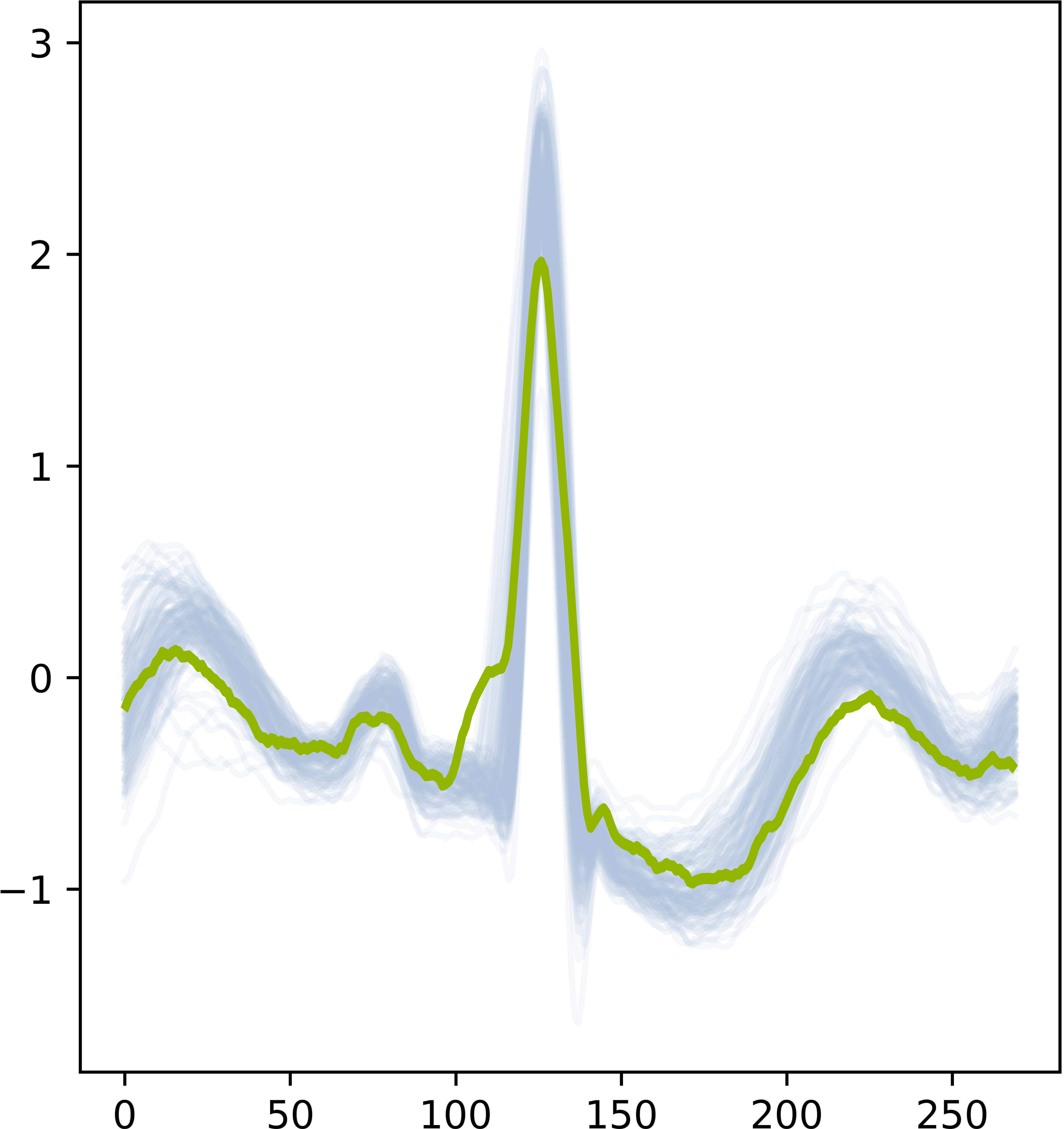} &
\includegraphics[width=0.19\textwidth]{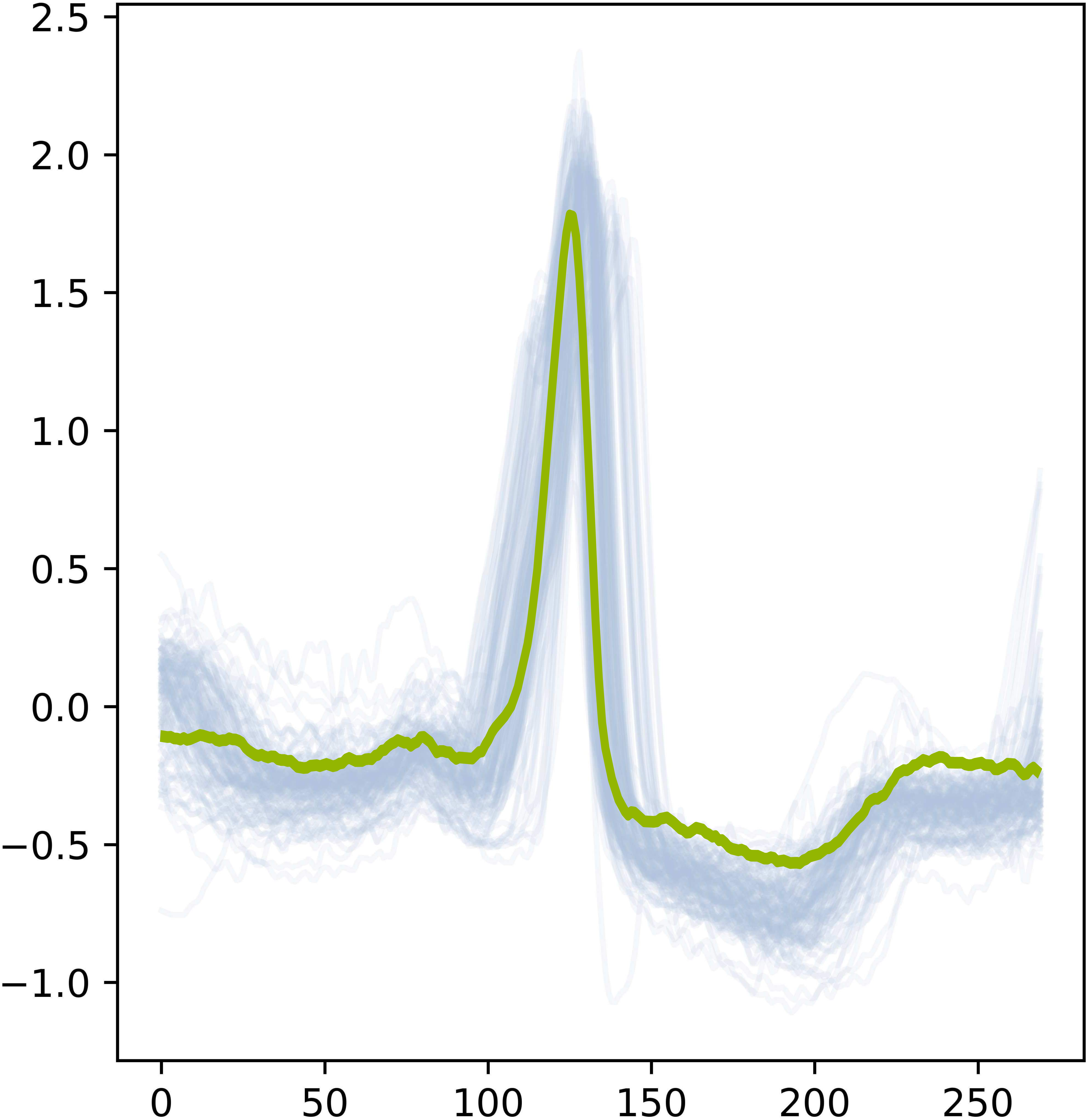} & 
\includegraphics[width=0.19\textwidth]{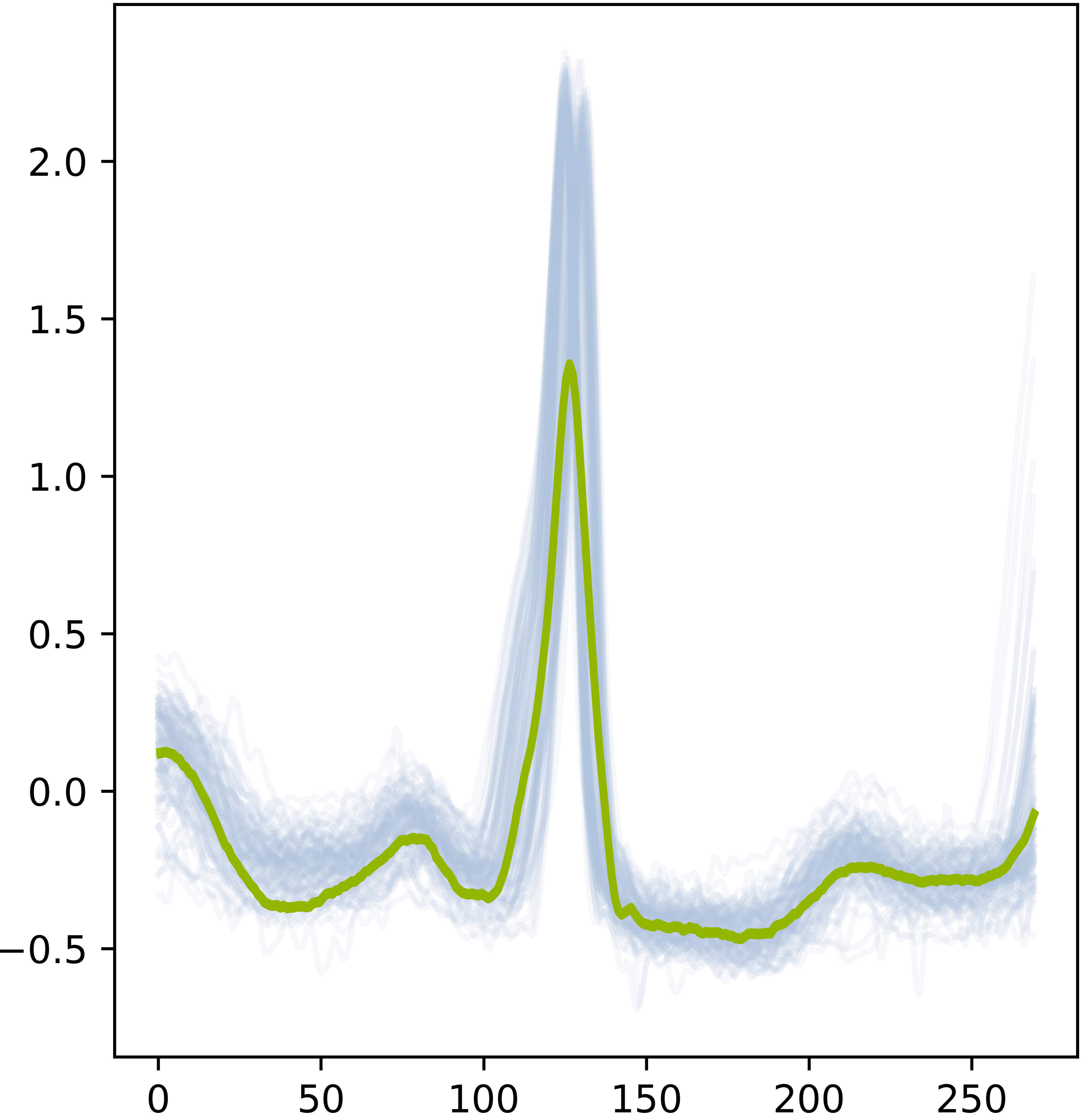} &
\includegraphics[width=0.186\textwidth]{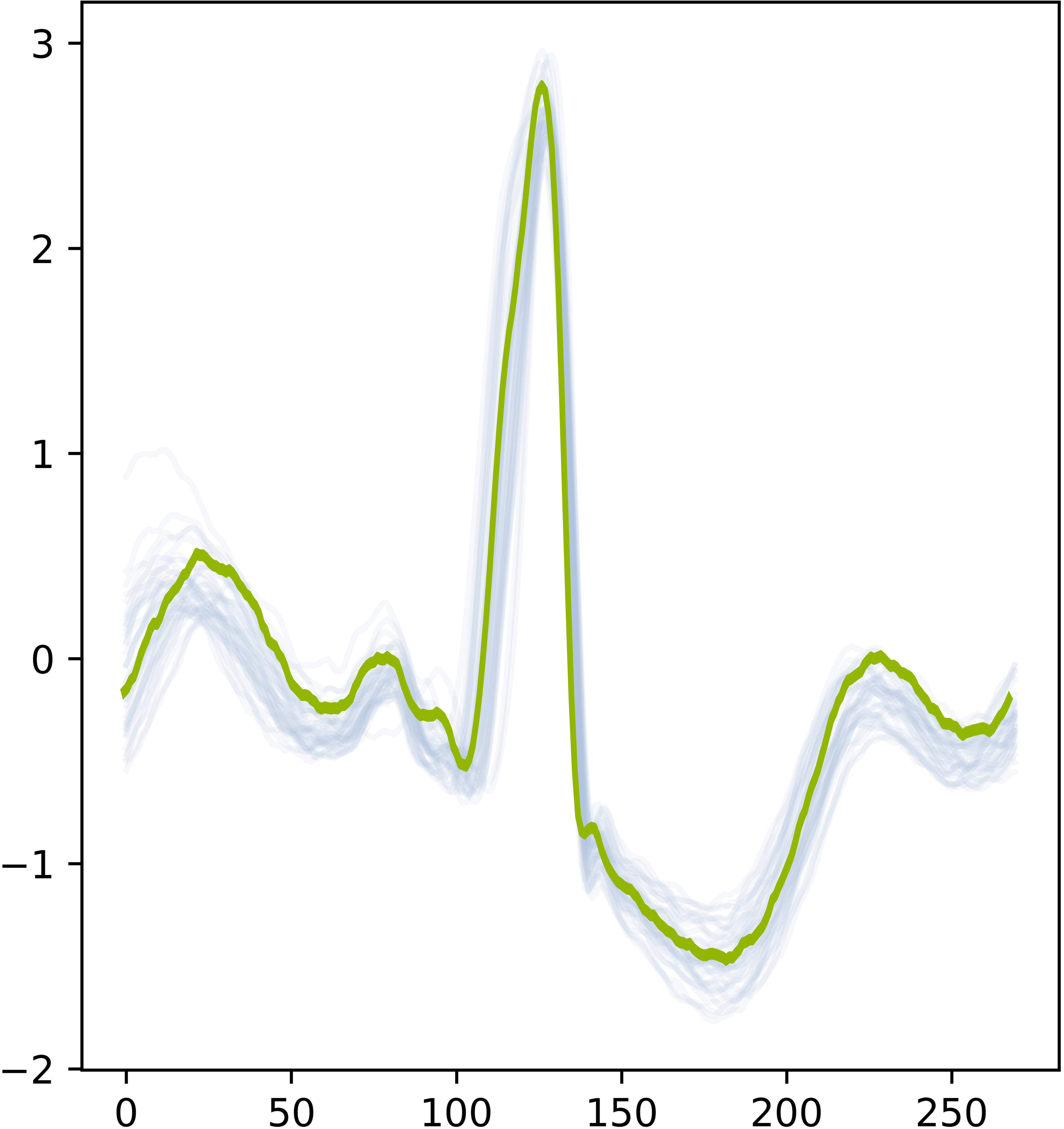} &
\includegraphics[width=0.19\textwidth]{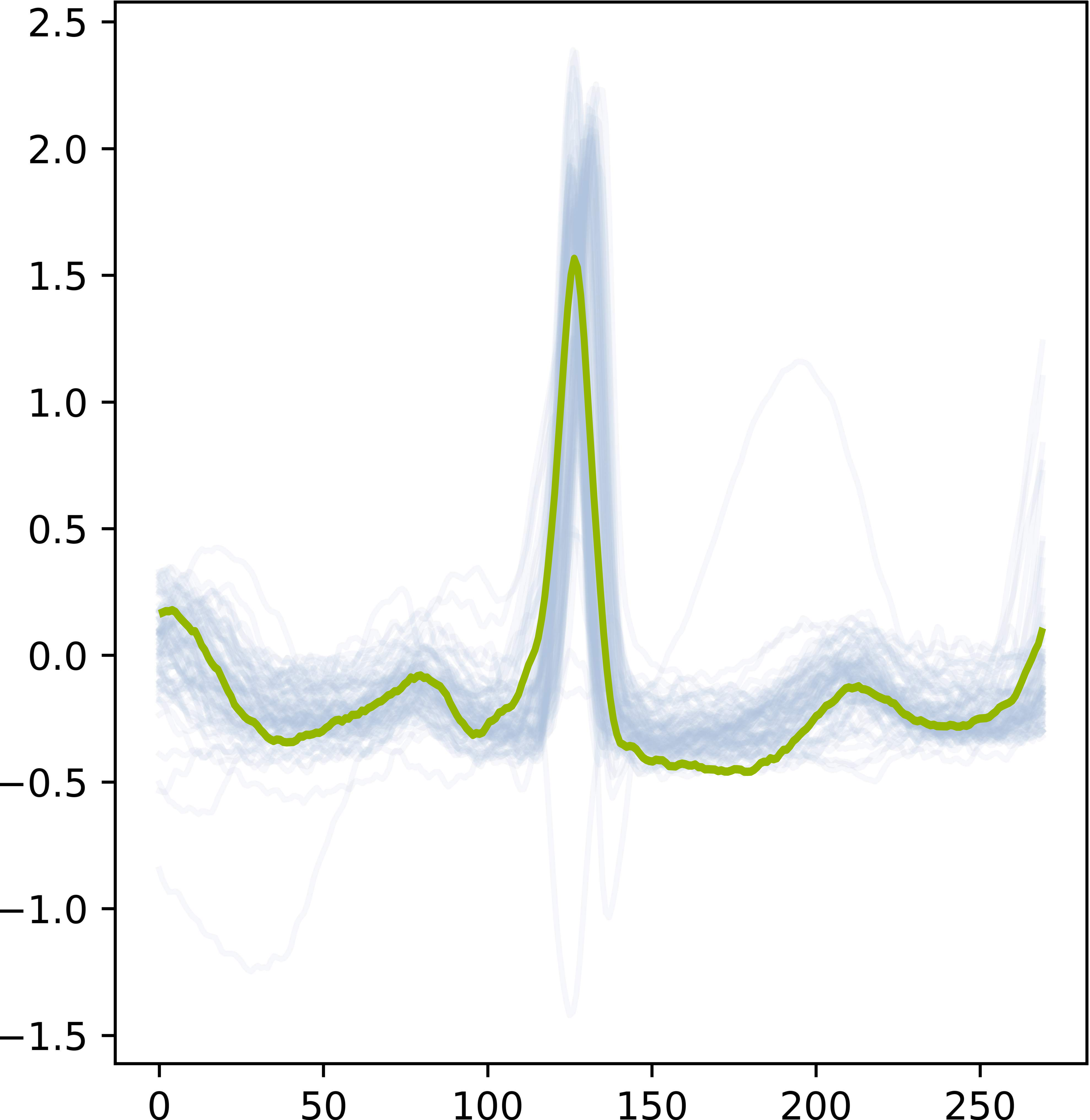} 
\end{tabular}
\caption{Examples of synthetic heartbeats for classes N,V and F. The blue background represents different clusters of the real dataset, while the green depicts synthetic heartbeats.}
\label{fig:fake_real_2}
% \end{figure*}
\end{sidewaysfigure}

% \begin{sidewaysfigure}[]
% \centering
% \begin{tabular}{cccccc}
% & cluster 1 & cluster 2  & cluster  3 &  cluster 4 & cluster 5  \\
% \multirow{3}{*}[15.9ex]{\rotatebox{90}{Class N}} & 
% \includegraphics[width=\figsize\textwidth]{N1.pdf} &
% \includegraphics[width=\figsize\textwidth]{N2.pdf} & 
% \includegraphics[width=\figsize\textwidth]{N3.pdf} &
% \includegraphics[width=\figsize\textwidth]{N4.pdf} &
% \includegraphics[width=0.185\textwidth]{N5.pdf} \\
% \multirow{3}{*}[15.9ex]{\rotatebox{90}{Class V}} & 
% \includegraphics[width=0.185\textwidth]{V1.pdf} & 
% \includegraphics[width=\figsize\textwidth]{V2.pdf} & 
% \includegraphics[width=\figsize\textwidth]{V3.pdf} &
% \includegraphics[width=0.185\textwidth]{V4.pdf} & 
% \includegraphics[width=0.185\textwidth]{V5.pdf} \\
% \multirow{3}{*}[15.9ex]{\rotatebox{90}{Class F}} &
% \includegraphics[width=0.185\textwidth]{F1.pdf} &
% \includegraphics[width=\figsize\textwidth]{F2.pdf} & 
% \includegraphics[width=\figsize\textwidth]{F3.pdf} &
% \includegraphics[width=0.185\textwidth]{F4.pdf} &
% \includegraphics[width=0.185\textwidth]{F5.pdf} 
% \end{tabular}
% \caption{Examples of synthetic heartbeats for classes N,V and F. The blue background represents different clusters of the real dataset, while the green depicts synthetic heartbeats.}
% \label{fig:fake_real_2}
% % \end{figure*}
% \end{sidewaysfigure}

\begin{table}[b]
\caption{Recognition rate of real and fake heartbeats as real heartbeats for the classes (N, V, F) of  cardiologists 1, 2, and 3.}
\centering % centering table 
\label{medical_validation}
\begin{tabular}{ccccc}
\hline
\multicolumn{2}{c}{}                                              & Class N & Class V & Class F \\ \hline
\multicolumn{1}{c}{\multirow{2}{*}{Cardiologist 1}} & real as real & 94.1\%  & 85.7\%  & 83.3\%  \\ \cline{2-5} 
\multicolumn{1}{c}{}                               & fake as real & 100\%   & 100\%   & 100\%   \\ \hline
\multicolumn{1}{c}{\multirow{2}{*}{Cardiologist 2}} & real as real & 100\%   & 100\%   & 83.3\%  \\ \cline{2-5} 
\multicolumn{1}{c}{}                               & fake as real & 100\%   & 71.4\%  & 83.3\%  \\ \hline
\multicolumn{1}{c}{\multirow{2}{*}{Cardiologist 3}} & real as real & 100\%   & 100\%   & 83.3\%  \\ \cline{2-5} 
\multicolumn{1}{c}{}                               & fake as real & 100\%   & 100\%   & 83.3\%  \\ \hline
\end{tabular}
\end{table}

The quality of the generated ECG heartbeats was further evaluated in a blind qualitative assessment by three cardiologists. The assessment consists of presenting shuffled examples of real and synthetic cardiac cycles of each class to the cardiologists and verifying if these examples correspond to their classes. For this purpose, we randomly chose a set of 30 samples of real heartbeats from the used training dataset (N, V, and F classes) and a set of 30 heartbeats generated by our approach (N, V, and F classes).
The three cardiologists successfully identified the classes of the presented cardiac cycles. First, we calculated the recognition rate of real and fake heartbeats as real, disregarding their categories. The recognition rate of real heartbeats as real represents the rate achieved by a cardiologist in validating that real heartbeats belong to their corresponding classes. The recognition rate of fake heartbeats as real corresponds to the validation rate that fake heartbeats belong to their classes. The first cardiologist validated 90\% of the real heartbeats as real ones and 100\% of the fake heartbeats as real ones. The second cardiologist achieved 96.6\% and 90\% as the recognition rates for real and fake heartbeats as real ones, respectively. The third cardiologist identified 96.6\% of both real heartbeats as real heartbeats and fake heartbeats as real heartbeats. The obtained recognition rates of each class are detailed in Table \ref{medical_validation}. The three cardiologists achieved good rates for both real and fake cases with the different types of heartbeats. These results show the robustness of the proposed approach in generating ECG signals similar to real signals with realistic wave morphologies.

\subsubsection{Quantitative Evaluation}
\paragraph{ECG classification:}
The quantitative evaluation is first carried out by training three state-of-the-art heartbeat classification baselines \cite{kachuee2018ecg,acharya2017deep,kumar2019arrhythmia} on our data following three settings:
\begin{itemize}

    \item Setting 1: The training is performed only with real training dataset without adding synthesized ECG signals.
    
    \item Setting 2: The MIT-BIH arrhythmia real dataset is augmented by synthetic ECG signals generated using a standard GAN  \cite{goodfellow2014generative} taking as its input noise vectors and class embeddings. The generator network is composed of a succession of FC and LeakyReLU layers, and finally outputs an entire ECG heartbeat signal.
 
    \item Setting 3: Same as setting 2, but the fake signals are generated using a stat-of-art advanced GAN \cite{nour2021Disentangling}. 
    
     \item Setting 4: Same as setting 2, but the MIT-BIH arrhythmia real dataset is augmented by synthetic ECG signals obtained by our generation approach.
\end{itemize}

These settings lead to 12 trained models that are then validated on the same test set consisting on real ECG signals randomly selected from the MIT-BIH arrhythmia dataset unseen during the training phases of the GAN and classifiers. The idea behind considering settings 1 and 4 is to highlight the contribution of data augmentation using our approach in improving ECG heartbeat classifiers, while the experiments related to settings 2 and 3 demonstrate that our generation approach outperforms the standard GAN and the state-of-the-art advanced GAN \cite{nour2021Disentangling} for the purpose of ECG signal generation.

Before discussing the obtained results, we introduce the competing classification baselines. The classifier presented by Kachuee \etal \cite{kachuee2018ecg} is composed of a 1-D convolutional layer fulfilled by five residual convolution sets, two FC layers and a softmax layer to generate the class probabilities. Every residual set is made up of two 1-D convolution layers, two ReLU activation layers, a residual skip connection, and finally a pooling layer.
The classifier architecture presented by Acharya \etal \cite{acharya2017deep} consists of three 1-D convolution layers, each followed by a max-pooling layer. Then, three FC layers are applied to the output data with a softmax function at the end of the last layer. 
The classification model developed by Kumar \etal \cite{kumar2019arrhythmia} is made up of four blocks, including a FC layer succeeded by a batch normalization layer and ReLU activation function. Finally, a FC layer and a softmax activation function are performed on the output of the last block. 

% \begin{figure*}[t]
% \centering
% \includegraphics[width=1\textwidth]{classifier1_kachuee.pdf}
% \caption{Classification results for \cite{kachuee2018ecg}.}
% \label{classfier1}
% \end{figure*}

% \begin{figure*}[t]
% \centering
% \includegraphics[width=1\textwidth]{classifier2_achrya.pdf}
% \caption{Classification results for \cite{acharya2017deep}.}
% \label{classfier2}
% \end{figure*}

% \begin{figure*}[t]
% \centering
% \includegraphics[width=1\textwidth]{classifier3_kumar.pdf}
% \caption{Classification results for \cite{kumar2019arrhythmia}.}
% \label{classfier3}
% \end{figure*}

\begin{figure*}[th!]
\centering
\includegraphics[width=0.9\textwidth]{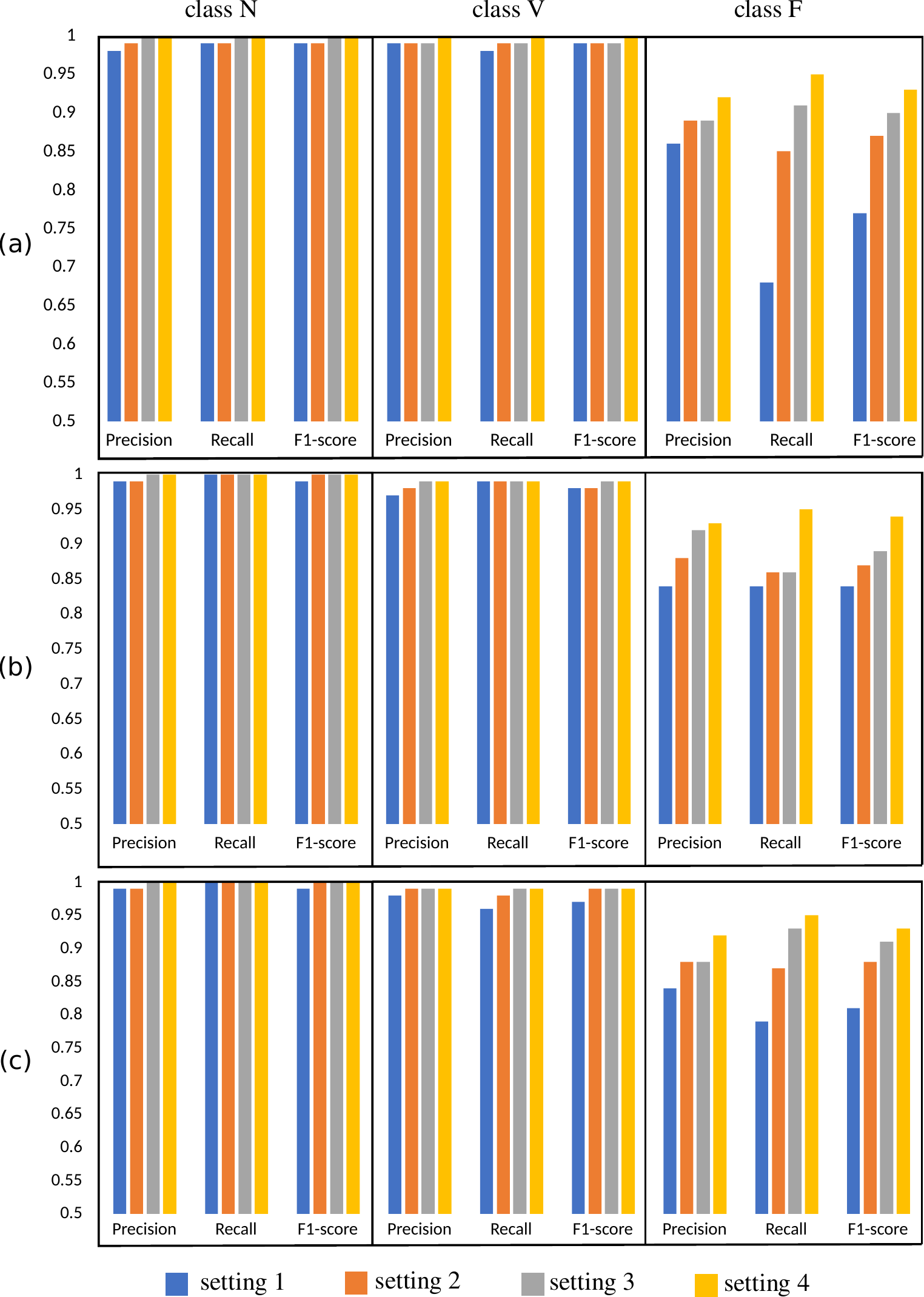}
\caption{Performance results of the three classification baselines \cite{kachuee2018ecg, acharya2017deep, kumar2019arrhythmia} on the classes N, V, and F. Each line (a), (b), and (c) represents the detailed performance of the baselines for the respective classes across the four settings.}
\label{fig:classfier_grouped}
\end{figure*}

Figure \ref{fig:classfier_grouped} summarize the classification results obtained for the three classes of heartbeats. The presented results demonstrate that using fake ECG heartbeats synthesized by our method as additional data with the training set of the real dataset improved the classification performance (\ie precision, recall, and F1 score). Particularly, they achieve a significantly higher performance when adding synthetic ECG heartbeats of the minority class F to the training dataset. As an example, the classifier \cite{kachuee2018ecg} achieved higher performance metrics in setting 4 compared to setting 1 for class F. The precision, recall, and F1 score were (0.92, 0.95, and 0.93) in setting 4, whereas they were (0.86, 0.68, and 0.77) in setting 1. The performance of the two other classes is faintly improved. For the same classifier, the performance has been improved by (2\%, 1\%, and 1\%) and (1\%, 1\%, and 1\%) from setting 1 to 4 in classes (N and V) respectively. Furthermore, we observe that classification models trained with added fake ECG heartbeats generated using our approach significantly outperform classification models trained with added fake ECG heartbeats generated by the standard GAN approach, with higher precision, recall, and F1-score values. It is also clear that the performance of three classifiers in setting 4 is better than the classifiers performance in setting 3 where synthetic data from the stat-of-art advanced GAN model are used.

The classifier model \cite{acharya2017deep} achieved in class F classification (0.93, 0.95, and 0.94) for (precision, recall, and F1 score) when using synthetic data from obtained by our approach in setting 4 while classifiers in settings 2 and 3 obtained (0.88, 0.86, and 0.87) and (0.92, 0.86, and 0.89), respectively. 

These results show the robustness of our approach in generating higher quality and more realistic ECG signals compared to state-of-the-art GAN approaches, including both standard and advanced methods. This clearly indicates that incorporating statistical shape modeling and prior knowledge of ECG patterns significantly enhances the generation process, resulting in more realistic signals.
%than a standard GAN approach \addNN{and the stat-of-art advanced GAN.} 

\paragraph{Evaluation metrics:}
Several metrics were considered to assess the quality of the generated signals: Root Mean Squared Error (RMSE), Mean Absolute Error (MAE), Mean Squared Error (MSE), Earth Mover’s Distance (EMD), Dynamic time warping (DTW). The obtained values for our generation method and other GAN models are shown in Table \ref{metrics}} where sets of 500 real and synthetic samples were used. We can observe that our generation approach  outperforms the other generation baselines across all heartbeats classes for the different metrics which demonstrate that our model is more effective in generating ECGs compared with other approaches. For instance, we achieved (2.64e-3, 7.0.2e-6) of (RMSE, MSE) in class N, while the standard GAN and the stat-of-art advanced GAN achieved (4.53e-3, 2.05e-5) and (3.92e-3, 1.54e-5), respectively.

\begin{table*}[b]
\caption{Obtained evaluation metrics for our approach and state-of-the-art GAN approaches .}
\vspace*{2mm}

\label{metrics}
\resizebox{\textwidth}{!}{
\begin{tabular}{cccccccccc}
\hline
\multirow{2}{*}{} & \multicolumn{3}{c}{Class N}                                            & \multicolumn{3}{c}{Class V}                                            & \multicolumn{3}{c}{Class F}                                            \\ \cline{2-10} 
                  & \multicolumn{1}{c}{\cite{goodfellow2014generative}} & \multicolumn{1}{c}{\cite{nour2021Disentangling}} & Ours    & \multicolumn{1}{c}{\cite{goodfellow2014generative}} & \multicolumn{1}{c}{\cite{nour2021Disentangling}} & Ours    & \multicolumn{1}{c}{\cite{goodfellow2014generative}} & \multicolumn{1}{c}{\cite{nour2021Disentangling}} & Ours    \\ \hline
RMSE              & \multicolumn{1}{c}{$4.53\mathrm{e}{-3}$}  & \multicolumn{1}{c}{3.92e-3}  & 2.64e-3 & \multicolumn{1}{c}{3.65e-3}  & \multicolumn{1}{c}{3.19e-3}  & 2.78e-3 & \multicolumn{1}{c}{2.38e-3}  & \multicolumn{1}{c}{2.34e-3}  & 2.26e-3 \\ \hline
MAE               & \multicolumn{1}{c}{3.52e-3}  & \multicolumn{1}{c}{2.89e-3}  & 2.28e-3 & \multicolumn{1}{c}{2.38e-3}  & \multicolumn{1}{c}{2.37e-3}  & 2.14e-3 & \multicolumn{1}{c}{1.82e-3}  & \multicolumn{1}{c}{1.16e-3}  & 1.16e-3 \\ \hline
MSE               & \multicolumn{1}{c}{2.05e-5}  & \multicolumn{1}{c}{1.54e-5}  & 7.02e-6 & \multicolumn{1}{c}{1.33e-5}  & \multicolumn{1}{c}{1.02e-5}  & 7.76e-6 & \multicolumn{1}{c}{5.7e-6}   & \multicolumn{1}{c}{5.5e-6}   & 5.13e-6 \\ \hline
EMD               & \multicolumn{1}{c}{7.32e-3}  & \multicolumn{1}{c}{5.04e-2}  & 3.76e-2 & \multicolumn{1}{c}{4.29e-2} & \multicolumn{1}{c}{3.81e-2}  & 3.25e-2 & \multicolumn{1}{c}{3.6e-2}   & \multicolumn{1}{c}{2.78e-2}  & 1.84e-2 \\ \hline
DTW               & \multicolumn{1}{c}{7.44}     & \multicolumn{1}{c}{5.75}     & 4.22    & \multicolumn{1}{c}{23.12}    & \multicolumn{1}{c}{22.43}    & 22.27   & \multicolumn{1}{c}{18.95}    & \multicolumn{1}{c}{17.19}    & 15.82   \\ \hline
\end{tabular}
}
\end{table*}
%======================================================================
\section{Conclusion\label{sec:conclusion}}
In this article, we proposed a novel GAN method for generating ECG signals. The proposed method leverages statistical shape modeling of the ECG signal dynamics to integrate prior knowledge of ECG patterns into the deep generation process. Furthermore, we proposed modeling ECG signals as a 2-D pattern representing its temporal and amplitude dynamics in order to generate realistic signals. The experimental results obtained on the MIT-BIH arrhythmia database showed the effectiveness and robustness of the proposed approach in synthesizing faithful ECG signals. Indeed, the synthetic ECG signals include a realistic ECG morphology and capture similar dynamics as real signals. Moreover, we demonstrated that our method outperforms other stat-of-art ECG generation methods and can improve the performance of known arrhythmia classification algorithms.
In future work, we plan to generalize our generation approach by synthesizing other classes of arrhythmia (supraventricular ectopic beat and fusion beat) and other physiological time series data characterized by specific patterns, such as photoplethysmographic (PPG) signals. During the clustering step, the number of clusters was optimized using a brute force approach, involving exhaustive iterations. However, future research should explore more efficient optimization methods to determine the optimal number of clusters for achieving the best variations modeling.

\section*{Acknowledgment}
This work is supported by the German Academic Exchange Service (DAAD) (Transformation Partnership: Theralytics Project). 
We thank our cardiologists for their active participation in the evaluation of this work.

%\section*{References}
\bibliography{neurocomputing}

\end{document}